\documentstyle[12pt]{article}
\begin{document}

\font\bbb = msbm10
\def\Bbb#1{\hbox{\bbb #1}}
\font \lbb = msbm7
\def\LBbb#1{\hbox{\lbbb #1}}
\newcommand{\BB} {{\Bbb B}}
\newcommand{\CC}   {{\Bbb C}}
\newcommand{\RR}  {{\Bbb R}}
\newcommand{\ZZ} {{\Bbb Z}}
\newcommand{\EE} {{\Bbb E}}
\newcommand{\NN} {{\Bbb N}}

\baselineskip 22pt plus 2pt

\begin{center}
{\bf\large Berry and Pancharatnam Topological Phases of Atomic and Optical Systems}\\ \ \\
{\bf Y. Ben-Aryeh}\\
Physics Department\\
Technion--Israel Institute of Technology, Haifa 32000, Israel\\
e-mail: phr65yb@physics.technion.ac.il; Fax: 972-4-8295755\\ \ \\
\end{center}

\noindent{\bf Abstract} Theoretical and experimental studies of
Berry and Pancharatnam phases are reviewed.  Basic elements of
differential geometry are presented for understanding the
topological nature of these phases. The basic theory analyzed by
Berry in relation to magnetic monopoles is presented. The theory
is generalized to nonadiabatic processes and to noncyclic
Pancharatnam phases. Different systems are discussed including
polarization optics, $n$-level atomic system, neutron
interferometry and molecular topological phases. \pagebreak

\section{Introduction}
There is a large interest in topological phases which lead to
interesting physical phenomena in quantum-mechanics and optics.
Although the appearance of geometrical phases was known quite a
long time ago from Aharonov-Bohm (AB) effect [1] and molecular
spectroscopy [2,3], the general context in which such phases
appear has been analyzed by Berry [4].

Berry [4] analyzed the problem of a quantum-mechanical state
developing adiabatically in time under a slowly varying
parameter-dependent Hamiltonian. He has shown that when the
parameters return to their initial values after traversing a
closed path, the wavefunction acquires a 'geometric' phase factor,
dependent on the path, in addition to the well-known 'dynamical'
phase factor $\exp\left[-\frac{i}{\hbar}\int Edt\right]$. The
geometry underlying Berry's phase has been reformulated in terms
of fibre-bundle theory [5-9] by Simon [10]. Aharonov and Anandan
[11] removed the adiabatic restriction and replaced the notion of
parameter space by the notion of projective space of rays in
Hilbert space. Generalizations of the Berry's phase analysis to
degenerate states have been analyzed by Wilczek and Zee [12].

Non-abelian gauge potentials were introduced in Yang-Mills theory
already in 1954 [13]. In this theory [14] a generalization of
electromagnetism was made in which the complex wavefunction of a
charged particle is replaced by a wavefunction with two
components, $\psi=\psi^i(x)$, $i=1,2$.

Following Wilczek and Zee article [12] many investigations have
been made on non-abelian topological phases [15-21] and this area
became of special interest due to the possibility to use
non-abelian topological phases in quantum computation [22-26].

Classical analogies to the quantum Berry's phase have been treated
by Hannay [27]. He has treated a classical Hamiltonian which
depends on parameters which are changed very slowly, then the
adiabatic theorem states that the action variable $I$ of the
motion is conserved. By treating the change in the angle variable,
holonomy has been made [27,28]. Application of the analysis to
generalized harmonic oscillators, with and without the adiabatic
approximation, have been analyzed [27-30]. Relations between
geometric quantum phases and angles have been also analysed
[31,32].

There are two basic kinds of topological phases which have been
studied experimentally for polarized light: a) The phase acquired
by a light wave when its polarization state undergoes a sequence
of unitary transformations. This phase is equal to {\em half the
solid angle} substended by the circuit traced by the state on the
{\em Poincare Sphere}. This case is similar to the doubly
degenerate case treated by Berry [4], as will be described in
Section (3.1). The similarity between the two cases including the
factor half follows from the isomorphism of the problem with
spin-$\frac{1}{2}$ system. b) Topological phases acquired by a
light wave when its {\em direction of propagation} is slowly
cycled along a closed path on the sphere of directions in real
space. This topological phase is equal to the {\em solid angle of
the circuit in the space of directions of the wavevector}. This
case is similar to the case treated by Berry [4] (See Section 3.1)
of a spinning particle in a magnetic field $\vec{B}$ which is
slowly varied. Jordan has shown [33] that the Berry phases for
spins or helicities is a general property of the spin states
rather than that of the Hamiltonian.

The fundamental article, which led to various experiments on
topological phases in momentum space was published by Chiao and Wu
in 1986 [34], and immediately after that it was realized
experimentally by Tomita and Chiao [35].

Many experimental and theoretical studies analysed noncyclic
topological phases by following Pancharatnam [36-37] definition of
phase. The use of Berry and Pancharatnam phases to noncyclic
evolution raises the problem of gauge invariance which have been
treated in various works [38-42]. It has been suggested to treat
open cycles by closing the end points of the open cycles by
geodesic lines [38,39]. Although this approach has some
interesting implications, it is possible to avoid the use of this
'trick' by following the kinematical approach [40,41].

Bhandari and Samuel [43], and Bhandari in series of papers
[44,45], have studied, experimentally and theoretically,
interference effects of polarized light. The mathematical
isomorphism between polarization of light and the two-state
quantum system has been exploited. It has been shown [43-45] that
for the appearance of topological phases the evolution of the
state needs to be neither unitary nor cyclic and can be
interrupted by measurements. Some new effects have been shown: a)
In contradiction to the U(1) group by which phases is defined only
module 2$\pi$, the phases obtained by the SU(2) group are
unbounded and can have arbitrary values if the Poincare sphere is
traversed more than once. b) Phase shifts defined by Pancharatnam
criterion can have discontinuous jumps and can change sign for
small variations in the parameters near singular points.
Singularities occur {\em around or through} points where the two
interfering beams become orthogonal and for which Pancharatnam
phase is undefined. The phase jumps are always related to the
geometric phases which have global properties in contrast to the
dynamical phases which are changing continuously.

The acquisition of a geometric phase in neutron spin rotation has
been verified experimentally by Bitter and Dubbers [46] for the
special case of adiabatic evolution. Nonadiabatic geometric phase
was observed by Suter et al. [47], using NMR techniques and also
it has been shown by Simon et al. [48] that evolving geometric
phase can introduce frequency shifts. The idea of obtaining
Lorentz-group Berry's phases [49] in squeezed light has been
suggested by Chiao and Jordan [50]. Lorentz-group Berry's phases
have been observed in polarization optics [51]. Geometric phase
observations have been made at the single photon level [52,53],
and in two photon interference experiments [54,55]. Berry's
topological phases have been measured by using optical fibre
interferometers [56,57].

There are many experimental and theoretical works studying the
geometric phases in neuron interferometry [58] including among
others the studies of Wagh and his colleagues [59-61]. In Section
3.4.3 we will analyze some of these experiments and in Section
(4.3) we will review some other works in this field.

Most of the observations of Berry's phases are related to SU(2)
and Lorentz groups. Geometric phases for SU(3) representations and
three-level quantum systems have been treated in some articles
[62-65].

There is much interest in the use of Berry's phases to quantum
computation [22-26]. There are different problems which have been
treated for the abelian phases, including: a) topological phases
for entangled states [66-68].  b) Generalization of topological
phases to mixed states and decoherence [69-72]. c) Geometric
quantum computation implemented in NMR [73-74]. The present
article emphasizes only basic properties of topological Berry's
and Pancharatnam phases. The present Review does not discuss the
literature on AB and quantum Hall effects, or other specific
problems (see, e.g., in the book Shapere and Wilczek [75]) in
order not to widen too much the scope of the present Review. We
would like, however, to refer here to an interesting discussion on
Berry's phase in quantum Hall effect given in the book of Yoshioka
[76]. Also of special interest in the present Review are the works
discussing the relations between topological phases and Riemannian
metric [40,77-81].

A deep understanding of topological phases can be achieved by the
use of fibre-bundle theories. Following this idea the present
Review is arranged as follows: In Chapter 2 basic elements of
topology are summarized. This chapter is divided into three
Sections: (3.1) Manifolds and Lie groups; (3.2) Cartan exterior
algebra; (3.3) Fibre-bundles and differential geometry. In Chapter
3 general theories of topological phases are discussed. This
chapter is divided into 4 sections: (3.1) Berry's phases for
Schrodinger equation under the adiabatic approximation; (3.2)
Generalization of Berry's phases to nonadiabatic processes and its
relation to fibre-bundle theory; (3.3) Description of Berry's
phases for $n$-level atomic system by the use of Kahler
potentials; (3.4) Generalization of Berry's phases to noncyclic
processes by the use of Pancharatnam phase.

In Chapter 4 we review some experimental measurements of Berry's
phases. This chapter is divided into 4 sections: (4.1) Topological
phases in molecular spectroscopy; (4.2) Topological phases in
polarization optics; (4.3) Geometric phases in neutron
interferometry; (4.4) Berry's phases related to Lorentz group. In
Chapter 5 we summarize our results and conclusions.

\section*{Chapter 2. \ Basic Theory of Topology}

We use in the present Review the Einstein summation convention in
which there is an implicit sum over each pair of repeated indices.
We use overbar to indicate the complex conjugate (The asterisk is
used later to denote cotangent space).

\addtocounter{section}{1}
\subsection{Manifolds and Lie groups}
Assuming that the reader has a basic knowledge of group theory we
discuss only certain relations between Lie groups and manifolds.

\subsubsection{Definition of a Manifold} 
A real (or complex) $n$ dimensional {\em manifold} $M^n$ looks
like Euclidean space $\RR^n$ (or $\CC^n$) around each point. We
introduce a certain `{\em neighborhood}' $O_i$ covering a part of
$M^n$, where each $O_i$ is a subspace of $\RR^n$ (or $\CC^n$). It
is possible to determine from a manifold of dimension $n$ another
manifold of dimension $n-1$ by taking the `{\em boundary}' of an
$n$ manifold. For example the boundary of a disc is a circle. We
denote the boundary of a manifold $M$ as $\partial M$. The
boundary of a boundary is always empty, $\partial\partial M = 0$.

Suppose we have covering $\{O_i\}$ of a manifold $M$ and a mapping
$\phi_i$ from $O_i$ to $\RR^n$, then in the overlapping $O_i\cap
O_j$ the transformation from the coordinate system $\phi_i$ to the
coordinate system $\phi_j$ is given by the transition function
$$\phi_{ij} = \phi_j \phi^{-1}_i \eqno(2.1)
$$

In equation (2.1) the coordinates $\RR^n$ of the neighborhood
$O_i$ are transformed back to the manifold and then the
overlapping neighborhood $O_j$ is projected to its coordinates in
$\RR^n$.

\subsubsection{Complex projective $n$-space $\CC P^n$} 

$\CC P^n$ is defined to be the space of complex lines through the
origin of $\CC^{n+1}$. Thus the point $z_0, z_1, \dots z_n$ is
equivalent to the point $\mu z_0, \mu z_1, \dots, \mu z_n$ for all
$\mu \in \CC - 0$. If $z_i\neq 0$ on this line, we may associate
to this line the coordinates $z_0/z_i, z_1/z_i \dots z_n/z_i$ with
$z_i/z_i$ omitted.

In quantum-mechanics $\CC P^n$ describes the {\em rays} of the
wavefunctions
$$ |\psi\rangle = z_0|0\rangle + z_1|1\rangle + \dots z_n|n\rangle
\eqno (2.2)$$ where $|0\rangle, |1\rangle \dots |n\rangle$ is an
orthogonal basis function. The coordinates $z_j/z_i \ (j\neq i,
z_i\neq 0$) describe the space of rays, e.g., in the solution of
Schrodinger equation for $n$-level atomic system.

\subsubsection{SO(3) Manifold} 

The rotations of a vector $\vec{r}(t)$ in $\RR^3$ space are
described by the SO(3) group. for a 1 parameter group with angular
velocity $\vec{\omega}$ we get
$$\left. \frac{d\vec{r}(t)}{dt}\right|_{t=t_0} = \vec{\omega} \times
\vec{r}(t_0) \eqno(2.3) $$ Starting from a vector in $\RR^3$ given
by (0,0,1), the development of this vector by the SO(3) group will
lead to a vector $(x,y,z)$ with a norm 1 $(x^2 + y^2 + z^2=1)$.
The tip of this vector is moving on the surface of a sphere which
is defined as the {\em Poincare sphere}. An important example
which is related to the SO(3) group is given by the {\em Two-Level
System} (TLS) described by the wavefunction
$$|\psi\rangle = C_1|\psi_1\rangle + C_2 | \psi_2\rangle \ ,
\eqno(2.4)$$ where $C_1$ and $C_2$ are complex numbers. The
components of the {\em Bloch vector} are given by
$$ r_1=\frac{C_1\bar{C}_2+\bar{C}_1C_2}{\sqrt{2}}; \
r_2=\frac{C_1\bar{C}_2 - \bar{C}_1C_2}{\sqrt{2}i}; \
r_3=\bar{C}_1C_1-\bar{C}_2C_2 \eqno(2.5)$$ $r_1$ and $r_2$ are
defined as the components of the complex dipole vector while $r_3$
is defined as the inversion. The interaction (without any losses)
between TLS and resonant (or nearly resonant) em field leads
(usually in a rotated system) to rotation of the Bloch vector on
the Poincare sphere.

\subsubsection{SU(2) Manifold}

SU(2) algebra consists of 2 $\times 2$ Hermitian matrices with
trace zero where the basis of these matrices is given by Pauli
matrices $\sigma_1,\sigma_2$ and $\sigma_3$. From the connection
between the Lie algebra of SU(2) and SO(3) one can find the
correspondence between these groups. Assuming, for example,
1-parameter subgroup of SU(2):
$$\exp\left[\left(\frac{\sigma_3}{2i}\right)\theta\right]=\left(\begin{array}{cc}
e^{-i\theta/2} & 0 \\ 0  & e^{i\theta/2}\end{array}\right),
\eqno(2.6)$$ it corresponds to the 1 parameter subgroup of
rotation of $\RR$ in SO(3)
$$\exp\left[E_3\theta\right]=\left(\begin{array}{ccc} \cos\theta &
-\sin\theta & 0\\\sin\theta & \cos\theta & 0 \\ 0 & 0 & 1
\end{array} \right)\eqno(2.7)
$$
where $$E_3=\left(\begin{array}{crc} 0 & -1 & 0\\1 & 0 & 0\\0 & 0
& 0
\end{array}  \right); \ E_1 = \left(\begin{array}{ccr} 0 & 0 & 0\\0 & 0 & -1\\0 & 1 & 0
\end{array}  \right); \ E_2 = \left(\begin{array}{rcc} 0 & 0 & 1\\0 & 0 & 0\\-1 & 0 & 0
\end{array}  \right) \eqno(2.8)$$
More generally the correspondence follows from the fact that the
commutation relations (CR) for $\frac{\sigma_1}{2i}$,
$\frac{\sigma_2}{2i}$ and $\frac{\sigma_3}{2i}$ are equal to the
corresponding CR of $E_1$, $E_2$ and $E_3$.

We find that for a change of $\theta$ in the range $0\leq \theta
\leq 2\pi$ equation (2.7) describes a closed rotation around the
$z$ axis. For the same change of $\theta$ the corresponding curve
for SU(2) described by equation (2.6) is not a closed curve as it
starts at $I$ and ends in $-I$. Only two full rotations of $SO(3)$
will correspond to a full rotation of SU(2). The change in sign of
a spin or wavefunction
$\left(\begin{array}{c}C_1\\C_2\end{array}\right)$ under a SU(2)
$2\pi$ rotation is an important observable effect (see Section
(3.1)) for spin-$\frac{1}{2}$ particles.

\subsubsection{Cosets of Lie Groups} 

Let $G$ be a Lie group with an algebra of $g$ elements and $H\in
G$ a subgroup. Denoting the algebra elements of $H$ by $e,
H_2\dots H_h$ we form the product $a e = a$, $a H_2, \dots a H_h$.
This aggregate of products is denoted simply as $aH$. The group
elements of $G$ can be given as a finite number $g/h$ of distinct
sets of $h$ elements $H, aH, bH \dots$. These sets of elements are
defined as the {\em left cosets} of $H$ in $G$, and the division
of the $g$ elements into these left cosets is defined as $G/H$. We
have shown that SU(2) is the {\em double covering} of SO(3), so
that SO(3) can be written as $SO(3)=SU(2)/Z_2$, where $Z_2$ is the
group with the two elements $\pm 1$.

\subsubsection{SU(1,1) Manifold} 

The generators $K_1, K_2, K_3=J_3$ of the SU(1,1) group obey the
CR[82]: $$\left[K_i,K_j\right] = i\tilde{\epsilon}_{ijk}K_k \ .
\eqno(2.9)$$ Here $\tilde{\epsilon}_{ijk}$ is the Ricci tensor of
a non-Euclidean space:
$$\tilde{\epsilon}_{ijk}=(-1)^{\delta_{k,3}} \epsilon_{ijk}
\left[\tilde{\epsilon}_{123}=-1, \ \tilde{\epsilon}_{231}=1, \
\tilde{\epsilon}_{312}=1\right], \eqno(2.10)$$ where
$\epsilon_{123}=\epsilon_{231}=\epsilon_{312}=1$ and $\delta$ is
the Kronecker symbol. $\epsilon$ with nonclycic transformations of
the indices 1,2,3 gives -1, and all other components of
$\epsilon_{ijk}$ (where two or three indices are equal) vanish. An
important example describing SU(1,1) manifold appears in the
treatment of squeezed em fields. For a single mode of the em field
with creation and annihilation operators $a^\dagger$ and $a$,
respectively, where $[a,a^\dagger]=1$, the generators $K_1, K_2$
and $K_3$ can be given as [50]:
$$K_1=-i(aa-a^\dagger a^\dagger)/4; \
K_2=(aa+a^\dagger a^\dagger)/4; \ K_3=J_3=\frac{aa^\dagger +
a^\dagger a}{4} \eqno(2.11)$$

Squeezed em field are produced by the unitary operator
$$S=\exp[ir(K_1\cos\theta + K_2\sin\theta)] \eqno(2.12)$$
operating on the Fock state $|n\rangle$ or the coherent state
$|\alpha\rangle$.

Let us consider a Hamiltonian given as [82]
$$H=\tau_1K_1+\tau_2K_2+\tau_3K_3 \ . \eqno(2.13)$$
The Bloch equations for $\vec{K}=(K_1, K_2, K_3)$ are given by
$$\frac{d\langle\vec{K}\rangle}{dt}=\vec{\tau}\tilde{\times}\langle\vec{K}\rangle
\eqno(2.14)$$ where the $SU(1,1)$ vector product $\tilde{\times}$
is defined as [82]
$$\vec{\tau}\tilde{\times}\langle\vec{K}\rangle =
\tilde{\epsilon}_{ijk} \tau_i \langle K_j\rangle \ . \eqno(2.15)$$
During the evolution $\langle\vec{K}\rangle \stackrel{\sim}{\cdot}
\langle\vec{K}\rangle = - \langle K_1\rangle^2 - \langle K_2
\rangle^2 + \langle K_3 \rangle^2$ is conserved and the vector
$\langle\vec{K}\rangle$ which starts from
$\langle\vec{K}(0)\rangle = (0,0,1)$ stays on the unit
hyperboloid, i.e. on the {\em Poincare hyperboloid}. Due to this
conservation law one can specify the vector
$\langle\vec{K}\rangle$ by its coordinates $\langle{K}_1\rangle,
\langle{K}_2\rangle$.

\subsection{Cartan Exterior Algebra}
\subsubsection{Tangent and Cotangent Space} 

Let us consider a point $p$ on $n$-dimensional manifold surface
$M$ with coordinates $x^i$. A basis for the {\em tangent vector
space} $T_p$ of $M$ at $p$ is given by $(\partial/\partial x^i)$.
For a differential $dx^i$ of a coordinate $x^i$ we use the
definition $$\langle d x^i \frac{\partial}{\partial x^j} \rangle =
\frac{\partial x^i}{\partial x^j} = \delta^i_j . \eqno(2.16) $$
For a vector $(v^j \partial/\partial x^j)$ belonging to the
tangent vector space $T_p$ we get $$\langle dx^i\sum_j
v^j\frac{\partial}{\partial x^j}\rangle = v^i \eqno(2.17)$$ $dx^i$
or a linear combination $\sum_i a_i dx^i$ is defined as a {\em
differential form}. The differential forms on the manifold at
point $p$ are defined as the {\em cotangent space} $T_p^*$.

We can generalize the basic element of $T_p(M)$ and $T_p^*(M)$ to
tensor fields over $M$ [5-9].

\subsubsection{Cartan's Wedge Product} 

The Cartan's product defined as {\em wedge product} or as {\em
exterior product} is the antisymmetric tensor product of cotangent
space basis elements. For example $$dx\wedge dy =
\frac{1}{2}(dx(\times)dy-dy(\times)dx) = - dy \wedge dx
\eqno(2.18)$$ where $(\times)$ denotes here an oriented
two-dimensional multiplication surface instead of ordinary
multiplication and $\wedge$ denotes exterior product. By
definition
$$dx \wedge dx = dy \wedge dy = 0 \ . \eqno(2.19)$$ The
differential elements $dx$ and $dy$ are {\em differential
1-forms}. The wedge product is a rule for constructing {\em
2-forms} out of {\em 1-forms}. A $j$-form on a manifold at a point
$p$ is given as
$$\omega^j(p)=f_{i_1,i_2 \dots i_j}(p) dx^{i_1} \wedge dx^{i_2}
\wedge \dots \wedge dx^{i_j} \eqno(2.20)$$ where the antisymmetric
tensor $f_{i_1, i_2 \dots i_j} (p)$ having $j$ indices is
contracted with the wedge product of $j$ differentials

\subsubsection{Exterior Derivative and Stokes Formula} 

Exterior derivative, takes the $j$ forms into $j+1$ forms
according to the rule $$ \begin{array}{l}d(f(x)) = \frac{\partial
f}{\partial x^i} dx^i, \\ d(f_j(x)dx^j) = \frac{\partial
f_j}{\partial x^i} d x^i \wedge dx^j, \\ d(f_{jk}(x) dx^j \wedge
dx^k) = \frac{\partial f_{jk}}{\partial x^i} dx^i \wedge dx^j
\wedge dx^k, \ {\rm etc}\end{array}  \ . \eqno(2.21) $$ {\em We
use the convention that new differential line element is always
inserted before any previously existing wedge product.}

The exterior derivative on any form $\omega_j$ gives zero when
applied twice $$dd\omega(p)=0 . \eqno(2.22)$$ The rule for
differentiating the wedge product of a $j$-form $\alpha_j$ and a
$q$-form $\beta_q$ is given by $$d(\alpha_j\wedge \beta_q) = d
\alpha_j \wedge \beta_q + (-1)^j \alpha_j\wedge d\beta_q \ .
\eqno(2.23)$$

In an oriented manifold two coordinate systems $\phi_i$ and
$\phi_j$ in the overlapping $O_i \cap O_j$ are related by positive
definite Jacobians.

If $M$ is a $j$-dimensional oriented manifold with a boundary
$\partial M$, then {\em Stokes theorem} says that for any
$(j-1)$-form $\omega_{j-1}$ $$\int_M d\omega_{j-1} =
\int_{\partial M} \omega_{j-1} \ . \eqno(2.24) $$ For further
explanations about this theorem and applications see Refs. 5-9.

\subsubsection{Homology and Cohomology} 

For a smooth connected manifold we define a {\em j-chain} $a_j$ as
a formal sum $\sum_i c_i N_i$ where the $N_i$ are smooth
$p$-dimensional oriented submanifolds of $M$. If the coefficient
$c_i$ are real (complex) then $a_j$ is a real (complex) $j$ chain;
if the coefficients $c_i$ are integers $a_j$ is an integer chain.

We define ${\partial}$ as the operation of taking the oriented
boundary, i.e., $\partial a_j=\sum_i c_i \partial N_i$ is a $j-1$
chain. The set of {\em cycles} is defined as the set of $a_k$
chains for which $\partial a_j=0$, i.e. cycles are chains with no
boundaries. {\em Boundaries} $B_j$ are defined as chains which can
be written as $B_j= \partial a_{j+1}$. The {\em simplical
homology} of $M$ is defined as $$H_j=Z_j/B_j \eqno(2.25)$$ $H_j$
is the set of equivalence classes of cycles $Z_j$ which differ
only by boundaries, i.e., $z'_j\simeq z_j$ $if z'_j-z_j=\partial
a_{j+1}$.

We have exterior differential forms on $M^n$:
$$\begin{array}{l} A^j:= {\rm all \ smooth} \ j{\rm -forms \ on} \
M\\[0.25cm]
Z^j:= {\rm all \ forms} \ \omega_j \ {\rm which \ are} \ {\em
closed,}
\ {\rm i.e., \ for \ which} \ d\omega_j = 0.\\[0.25cm]
B^j:= {\rm all \ forms \ which \ are} \ {\em exact,} \ {\rm i.e.,
\ where} \ \omega_j = d\alpha_{j-1}\end{array} \eqno(2.26)$$ de
Rham cohomology is defined as $$H^j = Z^j/B^j, \ {\rm i.e., closed
\ forms \ modulo \ exact \ forms} \eqno(2.27)$$
$$\omega_j\simeq \omega'_j \ {\em iff} \ \omega_j=\omega'_j+
d\alpha_{j-1} \ {\rm for \ some} \ \alpha_{j-1} \ . \eqno(2.28)$$
We define $$b_j= {\rm dim} \ H_j = {\rm dim} \ H^j \eqno(2.29)$$
as the $j$'th Betti number, i.e., it is equal to the dimension of
$H_j$ which is dual to $H^j$. The alternating sum of the Betti
numbers for a smooth manifold with $n$ dimensions is {\em Euler
characteristic} given as $$\chi(M) = \sum^n_{j=0} (-1)^j b_j \ .
\eqno(2.30)$$ One can find in the literature relations between
Euler characteristic and Gauss-Bonnet theorems [5-9] representing
integrations over all the manifold giving certain integers.

\subsubsection{Complex Manifolds} 

For a manifold with $n$ complex coordinates
$$z_k=x_k + i y_k \ (k=1,2 \dots n)\eqno(2.31)$$
$$\frac{\partial}{\partial z_k} =
\frac{1}{2}\left(\frac{\partial}{\partial
x_k}-i\frac{\partial}{\partial y_k}\right); \
\frac{\partial}{\partial\bar{z}_k} =
\frac{1}{2}\left(\frac{\partial}{\partial x_k} +
i\frac{\partial}{\partial y_k}\right)
\eqno(2.32)$$
$$dz_k=dx_k+idy_k; \ d\bar{z}_k=dx_k-idy_k \eqno(2.33)$$
A function $f$ of the manifold complex coordinates is {\em
holomorphic} if
$$df=\sum^n_{k=1} \frac{\partial f}{\partial z_k} dz_k = \partial
f \eqno(2.34)$$ and for which
$$\bar{\partial}f = \sum^n_{k=1} \frac{\partial f}{\partial
\bar{z}_k} d\bar{z}_k = 0 \ . \eqno(2.35)$$ The {\em complex
tangent and cotangent} spaces are defined as:
$$T_c(M)=\left\{\frac{\partial}{\partial z_k}\right\} \ ; \ \
\bar{T}_c=\left\{\frac{\partial}{\partial \bar{z}_k}\right\}
\eqno(2.36)$$
$$T^*_C(M)=\{d z_k\} \ ; \bar{T}^*_c=d\bar{z}_k \ . \eqno(2.37)$$
The theory of complex manifolds will be used later in connection
with Kahler manifolds.\\

\subsection{Fibre-Bundles and Differential Geometry}

In the present Section basic concepts of {\em differential
geometry} in relation to {\em fibre-bundle} theories are explained
[5-9].

\subsubsection{Fibre-Bundle Definition and Properties}

{\em Fibre-bundle} is defined by the following collection of
requirements:
\begin{enumerate}
\item[a.] We have a certain manifold $M$ which is defined as the
{\em basis space} $X$.
\item[b.] We have another manifold $F$ called the {\em fibre}
which is defined over the space $X$.
\item[c.] The total space $E$ includes both the basis space $X$
and the fibre $F$ and a {\em projection} $\pi$ of $E$ onto $X:\pi
E\rightarrow X$. A fibre-bundle $E$ over $M$ with a fibre $F$ is
covered with a set of {\em local neighborhoods} $\left\{
U_i\right\}$ where in each $U_i$ the bundle $E$ is described by
the product $U_i \times F$.
\item[d.] The bundle $E$ is specified by a set of {\em transition
functions} $\left\{\phi_{ij}\right\}$ which transform the fibre
manifolds between two neighborhoods $U_i \cap U_j$.
\item[e.] There is a set of open coordinate neighborhoods
$U_\alpha$ covering $X$ where for each $U_\alpha$
$$\pi^{-1} U_\alpha \rightarrow U_\alpha \times F \eqno(2.38)$$
Although the {\em local topology} of the bundle is trivial the
{\em global topology} of the {\em fibre-bundle} may be quite
complicated as a consequence of nontrivial transition functions.
\item[f.] If $x\in X$, i.e., if $x$ is a certain element in the
base space $X$, $\pi^{-1}(x)$ is the {\em fibre} $f$ over $x$,
i.e., $f$ is the fibre corresponding to $x$. An element of the
bundle $E$ can be written as $(x,f)$, $f\in F$, $x\in U_\alpha$
where $U_\alpha$ is a certain neighborhood of $X$.
\item[g.] If $(x,f)\in U_\alpha \times F$ and $(x',f')\in U_\beta \times F$
then $(x,f)\sim(x',f')$ if $x=x'$ and $\phi_{\alpha\beta}(x)f=f'$
where $\sim$ denotes equivalence relation.
\end{enumerate}
A {\em section s} of a fibre-bundle is given by a preferred point
$s(x)\in f(x)$ on each fibre corresponding to a point $x$ of the
base manifold $X\equiv M$.

\subsubsection{Connection on Riemannian Manifolds} 

For a Riemannian manifold and a local coordinate system $(U; u^i)$
the length $ds$ of an {\em infinitesimal tangent vector} is given
by
$$ds^2 = g_{ij} du^idu^j \eqno(2.39)$$
where $g_{ij}=g_{ji}$. For the unit sphere, which is a special
case of Riemannian manifold, $$ds^2=d\theta^2+\sin^2\theta d
\phi\eqno(2.40)$$ where $\theta$ and $\phi$ are the spherical
angles.

In each neighborhood $U$ of the basis manifold, the fibres which
are expressed as $\pi^{-1}U$, can be given by a linear combination
of the {\em local frame coordinates} $\left\{\hat{e}_1, \hat{e}_2
\dots \hat{e}_n\right\}$. A curve $\vec{x}(t)$ on the manifold
which is a function of parameter $t$ can be {\em lifted} to a {\em
local} section in the fibre given as
$$s(\vec{x}) = \sum^n_{i=1} \hat{e}_i(\vec{x}) z^i(\vec{x})
\eqno(2.41)$$ The local basis of coordinates of the tangent space
$T(E)$ is given by $\left(\partial/\partial x^\mu, \
\partial/\partial z^i\right)$ and of the cotangent space is given
by $\left(dx^\mu, \ d z^i\right)$, where $x^\mu$ and $z^i$ are the
coordinate of the basis and the fibre, respectively. A curve
$\vec{x}(t)$ in the basis $M$ is {\em lifted} to a curve
$$C(t)=\left(x^\mu(t), z^i(t)\right) \eqno(2.42)$$
in the total bundle. Differentiation along $C(t)$ is given by
$$\frac{d}{dt} = \dot{x}^\mu\frac{\partial}{\partial x^\mu} +
\dot{z}^i\frac{\partial}{\partial z^i} \ , \eqno(2.43)$$ where the
transformation of the fibres obey the {\em parallel transport
equation}
$$\dot{z}^i + \Gamma^i_{\mu j} \dot{x}^\mu z^j=0 \eqno(2.44)$$
and $\Gamma^i_{\mu j}$ are referred to as the {\em Levi-Civita
connections} (or Christoffel symbols). We get
$$\frac{d}{dt} = \dot{x}^\mu\left(\frac{\partial}{\partial x^\mu}
- \Gamma^i_{\mu j} z^j\frac{\partial}{\partial z^i}\right) =
\dot{x}^\mu D_\mu \eqno(2.45)$$ where the operator in the brackets
of equation (2.45) is defined as the {\em covariant derivative}
$D_\mu$. One should notice that the covariant derivative assumes
the {\em parallel transport} of equation (2.44). The change in the
total space $E$ is split into {\em vertical} and {\em horizontal}
components. The vertical components with basis $(\partial/\partial
z^i)$ represent changes which are strictly in the fibre and are
{\em not coupled} to changes in the manifold basis. Such {\em
vertical} components have no effect on the physical state.
Eq.~(2.45), based on the paralel transport, describes the change
in the physical state along the curve $\vec{x}(t)$. The first term
in the brackets of this equation represents ordinary derivative
while the second term represents changes in the fibre which are
coupled to the manifold basis. The above equations have been used
for Schrodinger equation [83] and were reduced to simple forms
where the fibre was found to be the geometric Berry's phase. Under
a change in frame we obtain
$$z^{'i} = \phi_{ij}(\vec{x}) z^j \ , \ \ e'_j=e_i \phi^{-1}_{ij}
(\vec{x}) \eqno(2.46)$$ where these sections are invariant
$$s(\vec{x}) = e_iz^i=e'_iz^{'i}=s'(\vec{x}) \ . \eqno(2.47)$$
Curvature measures the extent to which parallel transport is path
dependent and it is given by the commutators of $D_\mu$ and
$D_\nu$. The calculations of the connections $\Gamma^i_{jk}$ and
the curvatures might be quite complicated for general cases and
the reader is referred to the literature [5-9]. Also one finds
equations relating the connections to the metric coefficients
$g_{ik}$. [5-9]

\subsubsection{Connections of Vector Bundles}

{\em Tangent bundle} $TM^n$ to a manifold is defined for each
patch of the manifold by a corresponding set of coordinates $u$
tangent to the manifold. the vector field $\vec{Q}$ tangent to the
manifold is given by (See Section 2.2.1)
$$\vec{Q}=\left(Q_1\frac{\partial}{\partial u^1}, \
Q_2\frac{\partial}{\partial u^2} \dots Q_n\frac{\partial}{\partial
u^n}\right) \eqno(2.48)$$ where $\frac{\partial}{\partial u^1}, \
\frac{\partial}{\partial u^2} \dots \frac{\partial}{\partial u^n}$
is the basis for the vector bundle and $Q_1, Q_2 \dots Q_n$ are
the components of the vector field. The {\em cotangent bundle}
$T^*M^n$ is a {\em covector} field where the basis for the
cotangent bundle is given by $du^1, du^2 \dots du^n$ and the
covector field components are given by $A_1, A_2, \dots , A_n$. In
the overlap between the two patches $U_1$ and $U_2$ the same
vector fields can be transformed from one patch $U_1$ to another
$U_2$ by transition functions [5-9]. The {\em connections} on a
surface $M^2$ in $\RR^3$ of a vector-bundle have special simple
forms. (A nice exercise which derives the Levi Civita connections
on the manifold $M^2$ can be found in Ref. 5.)

Let us assume that we have a manifold surface $M^2$ in the space
$\RR^3$ and let $\vec{v}$ be a vector field that is tangent to $M$
at a point $p$. Let us assume also that we have a curve on the
surface $M$ which is a function of {\em parameter} $t$ and whose
tangent at $p$ is the vector $\vec{x}$. We define the {\em
covariant differentiation} $\nabla_{\vec{x}} \vec{v}$ of a vector
field $\vec{v}$ at the point $p$ as the projection of the ordinary
derivative $\frac{d\vec{v}}{dt}$ on the vector $\vec{x}$. The
covariant differentiation is obtained by throwing away the normal
component of the ordinary partial derivative. The covariant
differentiation differs from the ordinary partial derivative and
the quantity that measures this difference is called the {\em
connection}. The components of the covariant differentiation of
the vector $\vec{v}$ can be written as [6]:
$$\left(\nabla_{\vec{x}} \vec{v}\right)^\alpha = \frac{d
v^\alpha}{d u^\beta} x^\beta + \Gamma^\alpha_{\beta\gamma}
v^\gamma x^\beta \  \eqno(2.49)$$ The first term on the right side
of Eq.(2.49) represents the ordinary derivative while the second
term represents the connection. We say that $\vec{v}(t)$ is {\em
parallel transported} along the curve which is a function of
parameter $t$ $$\nabla_{\dot{\vec{x}}} \vec{v}=0 \ , \eqno(2.50)$$
i.e., the covariant differentiation along the tangent
$\dot{\vec{x}}$  to the curve vanishes.

A geodesic curve is defined by
$$\nabla_{\dot{\vec{x}}}(\dot{\vec{x}}) = 0 \eqno(2.51)$$
so that the covariant differentiation of $\dot{\vec{x}}$ has only
a component normal to the surface, and the geodesic equation
obtains the form [5-9]:
$$\frac{d^2x^\mu}{dt^2} + \Gamma^\mu_{\lambda \sigma}
\frac{dx^\lambda}{dt} \frac{dx^\sigma}{dt} = 0 \eqno(2.52)$$ The
change in angle of the vector $\vec{v}$ when it is parallel
transported along a closed circuit on the manifold basis is called
the {\em holonomy angle}. The {\em holonomy matrix} is the matrix
which rotates the vector $\vec{v}$ for this closed parallel
transportation. For vector $\vec{v}$ in $\RR^3$ and a manifold
$M^2$ one gets a $3\times 3$ holonomy matrix. The set of all
holonomy matrices form a group called the {\em holonomy group}.

\subsubsection{Connections on Principal Bundles}

A {\em principal bundle} is a fibre-bundle where the fibre is a
Lie group $G$ acting on a manifold. Both the transition functions
and the fibres belong to $g$ algebra and act on $G$ by left
multiplication. The Maurer-Cartan form $g^{-1}dg$ is a matrix of
one-forms belonging to the Lie algebra $g$. The coordinates of the
principal bundle $P$ are given by $(x,g)$ where $g\in G$. A {\em
local section} of $P$ is a map from a neighborhood $U$ of the
manifold basis to $G$. Connection on a principal bundle provides a
rule for the parallel transport of sections.

In any matrix group $G$, $g^{-1} dg$ is a matrix with left
invariant 1-form entries. For example, in SO(2), for
$$g(\theta)=\left(\begin{array}{cc} \cos\theta & -\sin\theta\\ \sin\theta & \cos\theta
\end{array}\right) \eqno(2.53)$$ we have
$$g^{-1} dg = \left(\begin{array}{cc} 0 & -1\\1 & 0
\end{array}\right) (\times) d\theta \eqno(2.54)$$ and $d\theta$ is a rotation
invariant 1-form on the circle SO(2). Let $h$ be a given ({\em
fixed}) group element then $(hg)^{-1} d(hg)=g^{-1}h^{-1} h dg =
g^{-1} dg$, as claimed.

The connection on $P$ is 1-form $\omega$ in $T^*(P)$ whose
vertical component is the Maurer-Cartan form $g^{-1}dg$, and it
can be written as
$$\omega = g^{-1} Ag + g^{-1} dg \eqno(2.55)$$
where
$$A(\vec{x}) = A^a_\mu (\vec{x})\frac{\lambda_a}{2i} dx^\mu\eqno(2.56)$$ satisfy
$$\left[\frac{\lambda_a}{2i}, \ \frac{\lambda_b}{2i}\right] =
f_{abc}\left(\frac{\lambda_c}{2i}\right) \eqno(2.57)$$ where
$f_{abc}$ are the structure constants.

If $\vec{x}(t)$ is a curve in $M$, the sections $g_{ij}(t)$ are
defined to be parallel transported along $\vec{x}(t)$ if the
following matrix differential equation is satisfied
$$g^{-1}\frac{dg}{dt} + g^{-1} \left(A^a_\mu
(\vec{x})\left(\frac{\lambda_a}{2i}\right)
\frac{dx^\mu}{dt}\right) g = 0 . \eqno(2.58)$$ The curvature is
{\em Lie algebra matrix 2-form} given by [5-9]:
$$ \Omega = d\omega + \omega \wedge \omega = g^{-1} Fg
\eqno(2.59)$$ where
$$F=dA + A\wedge A \eqno(2.60)$$
Let us consider two overlapping neighborhoods $U$ and $U'$ and a
transition function $\phi$ relating the local fibre coordinates
$g$ and $g'$, respectively, in $U$ and $U'$, then [5-9]:
$$g'=\phi g \ . \eqno(2.61)$$ In the overlapping region $U \cap
U'$, $A$ transforms as
$$A'=\phi A \phi^{-1} + \phi d\phi^{-1} \eqno(2.62)$$
while $F$ transforms as
$$F'=\phi F \phi^{-1} \eqno(2.63)$$
The above equations for {\em Principal Bundles} are especially
important for the use of non-abelian geometric phases in holonomic
quantum computation [22-26].

\subsubsection{Chern Classes}

The use of {\em characteristic polynomial} has been developed
extensively in the literature [5-9] for various {\em classes}. We
give here short explanation about {\em Chern Classes}, in relation
to integration over manifolds [5-9].

Let $\alpha$ be a complex $k \times k$ matrix and $P(\alpha)$ a
polynomial in the components of $\alpha$. Then $P(\alpha)$ is
called {\em invariant polynomial} or {\em characteristic
polynomial} if
$$P(\alpha) = P(g^{-1} \alpha g)\eqno(2.64)$$
for all $g\in GL(k, \CC)$. An example of invariant polynomial is
given by Det$(I+\alpha)$ which is used to define the Chern
classes. The total {\em Chern form} is defined as:
$$C(\Omega) = {\rm Det}\left(I + \frac{i}{2\pi}\Omega\right) = C_0
+ C_1 (\Omega) + C_2(\Omega) + \dots \eqno(2.65)$$ where
$$C_0=1; \ C_1 = \frac{i}{2\pi} Tr(\Omega); \ C_2 =
\frac{1}{8\pi^2}\left\{Tr(\Omega \wedge \Omega)-Tr(\Omega)\wedge
Tr(\Omega)\right\}, \eqno(2.66)$$ etc.

If we integrate $C_j(\Omega)$ over any $2j$ cycle in $M$ with
integer coefficients, we obtain an integer! The {\em Chern
numbers} [5] of a bundle are the numbers which result from
integration characteristic polynomials over the entire manifold.
An example of Chern number will be described later (see Section
2.2.4 for the definitions of {\em cycles} and {\em cohomology
groups}, and see the relations between Chern numbers and
cohomology groups as analyzed in Refs. [5-9].)

\section*{Chapter 3. General Theories of Topological Phases}
\addtocounter{section}{1} \addtocounter{subsection}{-3}
\subsection{Berry's Phases for Schrodinger Equation under the
Adiabatic Approximation}

In the present Section we summarize the main results obtained by
Berry in his paper of 1984 [4]. Although there is no experimental
evidence for the existence of magnetic charges or monopoles, the
interest in such monopoles arose in the scientific area of Berry's
phases due to the fact that {\em formally} certain Berry's phases
systems have the same {\em mathematical  structure} as the Dirac
magnetic monopole [84-88].

\subsubsection{Magnetic Monopole}

We give here only a short description of the fibre-bundle
structure of magnetic monopole [Ref. 75, p. 119].

For a single monopole at the origin it is convenient to describe
the gauge potential by spherical coordinates.
$$A_r=A_\theta = 0, \ \  A_\phi=(\frac{n}{2})(1-\cos\theta)
\eqno(3.1)$$ $A_\phi$ has a `Dirac string' singularity along the
line $\theta=\pi$; there the magnitude of $A_\phi$
$$g^{\phi\phi} A_\phi A_\phi=\frac{n}{2}
\frac{(1-\cos\theta)^2}{\sin^2\theta} . \eqno(3.2)$$ The string
can be moved around by means of a gauge transformation
$$A \rightarrow A - \nabla \Lambda \eqno(3.3)$$ but it cannot be
removed.

In order to avoid singular gauge potential, Wu and Yang [87]
introduced the following construction. They covered $S^2$ with two
patches $S^+$ and $S^-$:
$$\begin{array}{ll} S^+ \equiv \left\{\theta, \phi; \ 0 \leq \theta <
\frac{\pi}{2} + \epsilon\right\}, \\ S^- \equiv \left\{ \theta,
\phi; \frac{\pi}{2} - \epsilon < \theta \leq \pi \right\},
\end{array} \eqno(3.4) $$
two open sets that respectively contain the northern and southern
hemispheres and whose intersection is an open set containing the
equator. Over each patch, the restricted bundle is isomorphic to
the trivial bundle $S^\pm \times U(1)$. We break $S^2$ into the
two hemispherical neighborhoods
$$\begin{array}{ll} S^+ \times U(1), \ {\rm coordinates} \ \theta,
\phi, e^{i\psi_+} \\ S^- \times U(1), \ {\rm coordinates} \
\theta, \phi, e^{i\psi_-} \ . \end{array} \eqno(3.5)$$ The
transition functions must be function of $\phi$ along $S^+\cap
S^-$ and must be elements of U(1) to give a {\em principal
bundle}. We therefore choose to relate the $S^+$ and $S^-$ fibre
coordinates as follows
$$e^{i\psi_-} = e^{i n \phi} e^{i\psi_+} \ . \eqno(3.6)$$
The {\em winding number $n$ must be integer} for the resulting
structure to be a manifold, i.e., the fibres must fit together
exactly when we complete a full rotation around the equator in
$\phi$. This is a topological version of the Dirac monopole {\em
quantization condition}.

Using the above construction we can specify a non-singular gauge
potential for the magnetic monopole field. One choice is
$$A_\phi^+ = (\frac{n}{2}) (1-\cos \theta) \ {\rm over} \ S^+; \
A_\phi^- = (\frac{n}{2}) (-1-\cos \theta) \ {\rm over} \ S^- \ .
\eqno(3.7) $$ After straightforward calculations one finds [5]
that the curvature of the magnetic monopole is given by
$$F=(\frac{n}{2}) \sin \theta d \theta \wedge d \phi \eqno (3.8) $$
The total Chern class of U(1) bundle is given by [5]
$$C(P) = 1 + C_1(P) = 1-\frac{F}{2\pi} \ . \eqno(3.9)$$
The integral of $C_1$ for the Dirac monopole U(1) bundle, over
$S^2$ is integer giving the monopole charge
$$\int_{S^2} \frac{F}{2\pi} = - n \ . \eqno(3.10)$$

\subsubsection{Berry's Treatment of Geometrical Phases Related to
Magnetic Monopole Formalism}

Due to the importance of Berry's article of 1984 [4], which led to
a breakthrough in this field, we summarize here some of its
essential results.

Using the adiabatic approximation Berry assumed physical systems
which are transported around a closed path in parameter space
$\vec{R}(t)$, with Hamiltonian $H(\vec{R}(t))$ such that
$\vec{R}(T) = \vec{R}(0)$, and for simplicity $\vec{R}$ was
assumed to be three dimensional. The path is called circuit and
denoted by $C$. In addition to the dynamical phase, a geometrical
phase is obtained. After some calculations and using Stokes
theorem a special form for the geometric phase was obtained [4]
$$\gamma_n(C) = - \int\int_C d \vec{S} \cdot \vec{V}_n (\vec{R})
\eqno(3.11)$$ where $d\vec{S}$ denotes area element in $\vec{R}$
space, $n$ denotes a certain nondegenerate eigenstate of the
Hamiltonian and $\vec{V}_n(\vec{R})$ is given as [4]
$$\vec{V}_n(\vec{R})\equiv Im \sum_{m\neq n} \frac{\langle
n(\vec{R})|\vec{\nabla}_{\vec{R}} H(\vec{R})|m(\vec{R})\rangle
\times \langle
m(\vec{R})|\vec{\nabla}_{\vec{R}}H(\vec{R})|n(\vec{R})\rangle}
{\left[E_m(\vec{R})-E_n(\vec{R})\right]^2} \ . \eqno(3.12)$$

Berry treated the case in which the circuit $C$ lies close to a
point $\vec{R}^*$ in parameter space at which the state $n$ is
{\em involved in degeneracy}. He considered the most common
situation, where the degeneracy involves only two states denoted
by $+$ and $-$, and where $E_+(\vec{R})\geq E_-(\vec{R})$. For a
circuit which starts at $+$ state, he obtained after some
calculations
$$V_+(\vec{R}) = \frac{\vec{R}}{2R^3}\eqno(3.13)$$
where by using equation (3.13) for $V_n(\vec{R})=V_+(\vec{R})$ in
equation (3.11), the phase change $\gamma_+(C)$ is equal to the
flux through $C$ of the magnetic field of a monopole with strength
$-\frac{1}{2}$ located at the degeneracy $\vec{R}=0$. Obviously
$V_-(\vec{R})=-V_+(\vec{R})$ so that $\gamma_-(C)=-\gamma_+(C)$.
The geometric phase factor associated with $C$ is then given by
$$\exp\left\{i\gamma_{\pm}(C)\right\} =
\exp\left[\pm\frac{1}{2}\Omega(C)\right] \eqno(3.14)$$ where
$\Omega(C)$ is the {\em solid angle} that $C$ substends at the
degeneracy.

Berry [4] treated the case of a particle with spin $S$ (integer or
half integer) interacting with a magnetic field $\vec{B}$ via the
Hamiltonian
$$H(\vec{B})=\kappa \hbar \vec{B}\cdot\vec{S} \eqno(3.15)$$ where $\kappa$
is a constant involving the gyromagnetic ratio and $\vec{S}$ is
the vector spin operator with $2S+1$ eigenvalues $n$ that lie
between $-S$ and $+S$. The eigenvalues are
$$E_n(\vec{B}) = \kappa \hbar B n\eqno(3.16)$$
and so there is a $2S+1$-fold degeneracy when $\vec{B}=0$. The
components of $\vec{B}$ correspond to the parameters $\vec{R}$
used in the previous analysis. The phase change $\gamma_n(C)$ has
been calculated [4] for the case in which $\vec{B}$ is slowly
varied (and hence the spin rotated) round a circuit $C$ in the
{\em direction} along $\vec{B}$.

After some calculations Berry [4] got the result:
$$\vec{V}_n(\vec{B})=\frac{n\vec{B}}{B^3}
\ . \eqno(3.17)$$ Now, the use of Eq. (3.13) shows that
$\gamma_n(C)$
 is the flux
through $C$ of the ``magnetic field'' of a monopole with strength
$-n$ located at the origin of magnetic field space. Thus the
geometric phase factor is given by
$$\exp\left\{i\gamma_n(C)\right\}=\exp\left\{-in \Omega
(C)\right\} \eqno(3.18) $$ where $\Omega(C)$ is the solid angle
that $C$ substends at $\vec{B}=0$.

\subsection{Geometric Description of the Berry's Phase without the
Adiabatic Approximation}

If a quantum system is initially in an eigenstate of the
Hamiltonian and is changing with time, according to slowly varying
parameters, the adiabatic conditions guarantee that it remains in
an eigenstate of the instantaneous Hamiltonian. Assuming an
initially eigenstate of the Hamiltonian
$$H(\vec{R})|n,\vec{R}\rangle=E_n(\vec{R})|n,\vec{R}\rangle,
\eqno(3.19)$$ and parameters $\vec{R}$, which are slowly varying
along a closed curve $C$ in parameter space in time $T$, the
geometrical phase factor is given by [4]:
$$\gamma_n(C)=i\oint
d\vec{R}\cdot\langle
n,\vec{R}|\vec{\nabla}_{\vec{R}}|n,\vec{R}\rangle \ .
\eqno(3.20)$$ This phase is given in addition to the dynamical
phase given by
$$\phi_{\rm dyn}=-\int^T_0 E_n(\tau)d\tau \ . \eqno(3.21)$$
Considering the state vector $|\psi(t)\rangle$, Aharanov and
Anandan [11] removed its dynamical phase factor, by defining a new
state:
$$|\phi(t)\rangle = \exp \left[ i\int^t_0
h(t')\right]|\psi(t)\rangle \ , \eqno(3.22)$$ where
$$h(t')=\langle\psi|\psi\rangle^{-1} {\rm Re}\langle
\psi(t')|H(t')|\psi(t')\rangle \ . \eqno(3.23)$$ The geometrical
phase factor is then given, without any adiabatic approximation,
by
$$\gamma_{\rm geom.} = i \oint d\vec{R}\cdot\langle \phi(\vec{R},
t)|\vec{\nabla}_{\vec{R}}|\phi(\vec{R},t)\rangle \ . \eqno(3.24)$$
The term $\langle
n,\vec{R}|\vec{\nabla}_{\vec{R}}|n,\vec{R}\rangle$ of equation
(3.20), or the term
$\langle\phi(\vec{R},t)|\vec{\nabla}_{\vec{R}}|\phi(\vec{R},t)\rangle$
of equation (3.24), is referred to as a vector potential.

Topological effects obtained from closed circuits are referred to
in the literature as holonomy (like the present treatment of  a
change of phase in Schrodinger equation, a change of angle of
polarization in optics, etc.). By using Stokes theorem,
integration over a closed circuit can be transformed to
integration over the closed surface. In topological theories
$d\vec{R}\cdot \langle n
\vec{R}|\vec{\nabla}_{\vec{R}}|n,\vec{R}\rangle$ can be considered
as `one form' while the integration over the surface is considered
as the integration over the `curvature two-form'.

As in electromagnetic theory the vector potential is defined up to
a gauge transformation. By performing the transformation
$$|n,\vec{R}(t)\rangle'=\exp(i\alpha_n(\vec{R}))|n,\vec{R}(t)\rangle
\eqno(3.25)$$ we induce a ``gauge transformation'' on
$\vec{A}_{\vec{R}}(n,t)$:
$$\vec{A}'_{\vec{R}}(n,t)=\vec{A}_{\vec{R}}(n,t)-\vec{\nabla}_{\vec{R}}(\alpha_n(\vec{R}))
\ . \eqno(3.26)$$ From the definition of Berry phase given by
equation (3.20) it is clear that the Berry phase is gauge
invariant for a closed circuit.

The same conclusion is obtained for the vector potential defined
by equation (3.24).

The movement of the wavefunction $\phi(\vec{R},t)$ as a function
of the parameters $\vec{R}$ is considered in fibre-bundle theories
as a movement along the basis $\phi(\vec{R},t)$. This movement can
be `lifted' to the `total space' as the `section' which includes
the change of the `fibre', i.e., the change in the geometric phase
as a function of the change of the `fibre-bundle basis'. The use
of equation (3.20) or (3.24) is explained in `fibre-bundle'
theories as parallel transport of the Schrodinger wavefunction
[83,64-65]. An important point here is that the change of the
fibre is given uniquely as a function of changes in the manifold
basis. For many experiments on geometric phases the total space is
given by the manifold SU(2) while the fibre is U(1) and the basis
is SU(2)/U(1). However, other manifolds can be used with different
fibres. In the above treatment both the dynamical and geometrical
phase are included in the fibre where the first part has local
properties (depending on the wavefunction at a certain time) while
the latter has global properties (depending on closed circuit in
parameter space).

The Berry's geometric phase is similar to the geometric phase
obtained in AB effect, in which electron acquires a topological
phase shift after encircling a solenoid. Although such phase shift
is calculated for a closed circuit [1], this phase can be observed
by a shift in the interference pattern for electrons propagating
from their source into two routes, to their final observation
points. Mathematically, therefore, a closed circuit is obtained by
reversing the propagation direction in one of the two electron
routes. In order to obtain the topological phase one has to
eliminate the dynamical contribution to the phase difference,
which can follow from different optical paths in the two routes
which interfere.

The advantage of using closed circuits for observing topological
phases is that the results are gauge invariant and also that
Stokes theorem can be used. The use of closed circuits shows the
relation between topological phase and singular points. For vector
bundles singularity is obtained at the point at which the vector
field, e.g., the vector potential vanishes, as in this point the
direction of the vector field is undefined. An important point in
this connection is that geometric phases, as those defined by
Berry [4,11] can be obtained only if the number of components of
$\vec{R}$ is larger than 1, so that it can encircle a singular
point. In Section 3.4 we generalize, however, the use of geometric
phases to open circle by following the definition of Pancharatnam
phase [36].

\subsection{Berry's Phases for a $n$-Level Atomic Systems and
K\"ahler Metric}
\subsubsection{Kahler Metric [5-9, 89-91]}

We define {\em complex exterior form} $\Lambda^{p,q}$ which have
basis containing $p$ factors of $dZ_k$  and $g$ factors of
$d\bar{Z}_j$ with corresponding wedge products [see the analysis
and explanations given in Section (2.2)]. The operators $\partial$
and $\bar{\partial}$ act as:
$$\begin{array}{l} \partial: \
\Lambda^{p,q}\rightarrow\Lambda^{p+1,q} \ , \\ \bar{\partial}: \
\Lambda^{p,q} \rightarrow \Lambda^{p,q+1}
\end{array} \eqno(3.27)$$ These operators satisfy the relations
$$\partial\partial\omega=0 \ , \ \
\bar{\partial}\bar{\partial}\omega=0 \ ,  \ \
\partial\bar{\partial}\omega=-\bar{\partial}\partial\omega
\eqno(3.28)$$ We have an {\em almost complex structure} if there
exists a linear map $J$ from $T(M)$ to $T(M)$ such that $J^2=-1$.
For example take a cartesian coordinate system $(x,y)$ on $\RR^2$
and define $J$ by the $2\times 2$ matrix
$$J\left(\begin{array}{c} x\\ y \end{array}\right)=
\left(\begin{array}{cc} 0 & -1\\ 1 & 0 \end{array}\right)
\left(\begin{array}{c} x\\ y \end{array}\right) \Rightarrow J^2
\left(\begin{array}{c} x\\ y \end{array}\right) = -
\left(\begin{array}{c} x\\ y \end{array}\right) \ . \eqno(3.29)$$
In this example $J$ is equivalent to multiplication by
$i=\sqrt{-1}$.

More generally operator $J$ is defined as an operator which have
eigenvalues $\pm i$. No $J$ can be found on odd-dimensional
manifold.

Let us consider a Hermitian metric on $M$ given by
$$ds^2 = g_{a\bar{b}} d Z_a d\bar{Z}_b \eqno(3.30)$$ where
$g_{a\bar{b}}$ is a Hermitian matrix. We define the {\em Kahler
form}
$$K=\left(\frac{i}{2} \right) g_{a\bar{b}} dZ_a \wedge d\bar{Z}_b \ .
\eqno(3.31)$$ It is easy to show that $K=\bar{K}$ [5].

A matrix is said to be a {\em K\"ahler metric} if $dK=0$, i.e., if
the K\"ahler form is {\em closed}.

The {\em Fubiny-Study metric} on $P_n(\CC)$ [see Section 2.1.2)]
is given by
$$ \begin{array}{ll} K  & =  \left(\frac{i}{2}\right)
\partial\bar{\partial}\ln\left(1+\sum^n_{\alpha=1} Z^\alpha\bar{Z}^\alpha\right)\\
&=\left(\frac{i}{2}\right)\left(\frac{dZ_\alpha\wedge
d\bar{Z}^\beta}{(1+\sum_\gamma Z_\gamma\bar{Z}^\gamma)^2}\right)
\left\{\left[\delta_{\alpha\beta}(1+\sum_\gamma
Z^\gamma\bar{Z}^\gamma)\right]-\bar{Z}_\alpha Z^\beta \right\}
\end{array} \ . \eqno (3.32)$$ This quite complicated equation has
a simple form for the standard metric on $S^2$ with radius
$\frac{1}{2}$ given in complex coordinates by
$$ds^2 = \frac{dx^2+dy^2}{1+x^2+y^2} = \frac{dZ
d\bar{Z}}{(1+Z\bar{Z})^2} \eqno(3.33)$$ and the K\"ahler form is
given by
$$K=\left(\frac{i}{2}\right)\frac{dZ \wedge
d\bar{Z}}{(1+Z\bar{Z})^2} = \frac{dx\wedge dy}{(1+x^2+y^2)^2} =
\left(\frac{i}{2}\right) \partial\bar{\partial}\ln(1+z\bar{z|}) \
. \eqno(3.34)$$ The ln terms in equations (3.32) and (3.34) are
referred to as {\em Kahler potentials}. The integration of (3.34),
multiplied by $\frac{1}{\pi}$, over all the manifold gives 1 (a
special form of Gauss-Bonnet theorem with Chern number 1).

\subsubsection{Berry's Phase for $n$-Level  Atomic System [92]}

Let us take a nondegenerate $n$-level atomic system so that the
quantum system of this Hilbert space is given by $n+1$ complex
amplitudes $Z^\alpha$ where superscripts of Greek indices range
from $0$ to $n$
$$|\psi\rangle = |Z^0, Z^1, \dots Z^n\rangle \eqno(3.35)$$
For a normalized state
$$\langle\psi|\psi\rangle=\delta_{\alpha\beta} \bar{Z}^\alpha
Z^\beta = \bar{Z}_\beta Z^\beta = 1 \eqno (3.36)$$ and this lies
on the unit sphere $S^{2n+1}$ in $\CC^{n+1}$.

For $Z^0\neq 0$ $P_n(\CC)$ may be given by complex coordinates
[see Section (2.1.2)]
$$W^i=Z^i/Z^0\eqno(3.37)$$ where Latin indices range from 1 to
$n$. Using units in which $\hbar=1$ and denoting time derivative
by an overdot one may write the evolution of the quantum state by
Hermitian Hamiltonian $H$ as
$$ |\dot{\psi}(t)\rangle = -i H(t)|\psi(t)\rangle \Rightarrow
\dot{Z}^\alpha = -i H^\alpha_\beta Z^\beta \ . \eqno(3.38)$$ If
the evolution undergoes a circuit in ray space, the original state
returns to itself up to a phase factor
$$|\psi(T)\rangle = e^{i\phi(T)}|\psi(0)\rangle \ . \eqno(3.39)$$
Part of this phase $\phi(T)$ may be identified as a dynamical
phase [see Section (3.2)]:
$$\epsilon(T)=-\int^T_0\frac{\langle\psi (t)|H(t)|\psi(t)\rangle}{\langle\psi(t)|\psi(t)\rangle}
\eqno(3.40)$$ but the remainder $$\gamma(T)=\phi(T)-\epsilon(T)
\eqno(3.41)$$ is geometrical and depends purely on the closed path
evolution of the ray in projective Hilbert space. After
straightforward calculations one gets [92]:
$$\dot{\gamma}(t)dt = \left(\frac{i}{2}\right)
\left(\frac{\bar{W}_idW^i-W_id\bar{W}^i}{1+\bar{W}_k W^k}\right) =
A \ , \eqno(3.42)$$ so that
$$\gamma(T)=\oint A \ , \eqno(3.43)$$ i.e., the Berry's phase is
obtained by integral of the one-form $A$ around the circuit in
$P_n(\CC)$. So that $A$ is the {\em connection for the geometrical
phase}.

By the {\em Stokes theorem}, one gets
$$\gamma(T)=\int\int_S F \eqno(3.44)$$ the integral, over a
surface $S$ bounded by the circuit of the {\em curvature 2-form}
$$F=dA=i\left\{\frac{\bar{W}_iW_j-(1+\bar{W}_kW^k)\delta_{ij}}{(1+\bar{W}_\ell
W^\ell)^2}\right\} dW^i\wedge d\bar{W}^j \ . \eqno(3.45)$$ Thus
$F$ is an explicit realization of the 2-form given by Simon [10]
in terms of the coordinates of $P_n(\CC)$.

\subsection{Pancharatnam Phase and Noncyclic Evolution}

Consider two normalized nonorthogonal Hilbert states $|A\rangle$
and $|B\rangle$, and assume further that $|A\rangle$ is exposed to
U(1) shift $e^{i\phi}$ [36]. The resulting interference pattern is
determined by
$$I=|e^{i\phi}|A\rangle + |B\rangle
|^2=2+2|\langle A|B \rangle|\cos[\phi-\arg \langle A|B\rangle]
\eqno(3.46)$$ where its maximum is obtained at the Pancharatnam
relative phase $\phi_0\equiv\arg \langle A|B\rangle$. This phase
is reduced to the U(1) case whenever
$|B\rangle=e^{i\theta}|A\rangle$ as it yields $\arg\langle
A|B\rangle = \theta$. In the original treatments of the Berry
phase [4] one considers a quantal system evolving around a closed
circuit, from an initial wavefunction $|A\rangle$ to a final
wavefunction $|B\rangle$ where $|B\rangle$ is obtained from
$|A\rangle$ be a cyclic evolution, i.e., by multiplication with a
U(1) phase factor. Although the initial phase of the quantal state
is defined arbitrarily, the phase difference between the state
$|A\rangle$ and $|B\rangle$ is well defined and can be observed by
interferometric methods. In a non-cyclic evolution if the final
wave $|B\rangle$ is superimposed on the initial state $|A\rangle$
only the component $\langle A|B\rangle|A\rangle$ along $|A\rangle$
interferes with $|A\rangle$. All other components of $|B\rangle$
which are orthogonal to $|A\rangle$ merely add to the intensity,
since their cross terms with $|A\rangle$ vanish. The {\em
`Pancharatnam connection'} defines the phase between $|B\rangle$
and $|A\rangle$ as $\phi_0=\arg\langle A|B\rangle$. The
interference amplitude $|\langle A|B\rangle$ differs from unity if
the evolution is non-cyclic [36] and this difference leads to
reduction of the visibility in the interference pattern. The
Pancharatnam phase is indeterminate if $|B\rangle$ is orthogonal
to $|A\rangle$. The use of Pancharatnam phases to noncyclic
evolution raises the problem of gauge invariance.

In order to treat this problem we introduce a more general
definition of the geometric phase which is defined also for open
circuits and is reduced to that described in the Sections
(3.1-3.3) for closed circuits. In Section (3.4.1) we develop such
definition and in Section (3.4.2) we apply it to neutron
interferometry.

\subsubsection{The Geometric Phase for Noncyclic Evolution}

Let us assume that under adiabatic approximation the initial state
$|n;\vec{R}(0)\rangle$ at time $t=0$ develop after time $t$ into
the state $|n,\vec{R}(t)\rangle$ where this development is not
cyclic, i.e. $\vec{R}(t) \neq \vec{R}(0)$. We define the geometric
phase as [93]:
$$\gamma_n(C) = \arg \langle n;\vec{R}(0)|n;\vec{R}(t)\rangle + i
\int^{\vec{R}(t)}_{\vec{R}(0)} d\vec{R}\cdot\langle
n,\vec{R}|\vec{\nabla}_{\vec{R}}|n,\vec{R}\rangle \eqno(3.47)$$
where now $C$ is not a closed circuit. This equation reduces to
equation (3.20) for a cyclic motion where the first term on the
right side of equation (3.47) vanishes. We have the following
relations:
$$R_e\langle n,\vec{R}|\vec{\nabla}_{\vec{R}}|n,\vec{R}\rangle = 0;
\langle n, \vec{R}|\vec{\nabla}_{\vec{R}}|n,\vec{R}\rangle=i
Im\langle n,\vec{R}|\vec{\nabla}_{\vec{R}}|n,\vec{R}\rangle \ .
\eqno(3.48)$$ Therefore we can write (3.47) as
$$\gamma_n(C) = \arg \langle n;\vec{R}(0)|n;\vec{R}(t)\rangle -Im
\int^{\vec{R}(t)}_{\vec{R}(0)} d\vec{R}\cdot\langle n
\vec{R}|\vec{\nabla}_{\vec{R}}|n,\vec{R}\rangle \ . \eqno(3.49)$$
The crucial point here is that $\gamma_n(C)$ is gauge invariant
since the change in the two terms on the right side of equation
(3.47) are equal in magnitude and opposite in sign. We can
generalize the definition of geometric phase for open cycles
without the adiabatic approximation as
$$\gamma_{\rm geom.}(C)=\arg
\langle\phi(\vec{R},0)|\phi(\vec{R},t)\rangle-Im
\int^{\vec{R}(t)}_{\vec{R}(0)} d \vec{R} \cdot
\langle\phi(\vec{R},t)|\vec{\nabla}_{\vec{R}}|\phi(\vec{R},t)\rangle
\eqno(3.50)$$ where the function $\phi(\vec{R},t)$ was defined in
Section (3.2). The above equations are obtained after eliminating
the dynamical phase from the time evolution and the definitions
given here coincide with the definitions of geometric phases of
Section (3.2) only for closed circuits.

If the wavefunction depends {\em only} on one parameter $S$ then
$\langle\psi(S)|\dot{\psi}(S)\rangle$ (where the derivative
$\dot{\psi}$ is according to parameter $S$) might be defined as a
dynamical phase [40]. For such cases we can define gauge invariant
phases for open circuits by subtracting this dynamical phase from
$\langle n,S(0)|n,S(t)\rangle$ [or
$\langle\phi(S(0))|\phi(S(t))\rangle$] [40]. We find that
different definitions of geometric phases can be used, but any
definition of geometric phase should be gauge invariant.

Samuel and Bhandari [39] and Jordan [38] solved the problem of
gauge invariant geometric phases for open circuits by closing the
end points of the open circuit by the shortest `geodesic' line.
Then the gauge invariant phase for the open circuit is equal to
that obtained from the closed circuit for which one can use the
definitions of Section (3.2). In order to explain this `trick' let
us use equation (3.47) for a geodesic line (for which
$\gamma_n(C)=0$), starting at $|n,\vec{R}(t)\rangle$ and ending at
$|n,\vec{R}(0)\rangle$. For this geodesic line we get
$\gamma_n({\rm geodesic})=0=\arg\langle
n,\vec{R}(t)|n,\vec{R}(0)\rangle + i\int^{\vec{R}(0)}_{\vec{R}(t)}
d\vec{R}_{\rm geod.} \cdot \langle
n,\vec{R}|\vec{\nabla}_{\vec{R}}|n,\vec{R}\rangle$ \\
so that $$\arg\langle
n,\vec{R}(0)|n,\vec{R}(t)\rangle=i\int^{\vec{R}(0)}_{\vec{R}(t)}d\vec{R}_{\rm
geod.} \cdot\langle
n,\vec{R}|\vec{\nabla}_{\vec{R}}|n,\vec{R}\rangle \eqno(3.51)$$
Substituting equation (3.51)in equation (3.47) we get
$$\gamma_n(C) = i \int^{\vec{R}(t)}_{\vec{R}(0)}
d\vec{R}\cdot\langle n,\vec{R}|\vec{\nabla}_R|n,\vec{R}\rangle +
i\int^{\vec{R}(0)}_{\vec{R}(t)} d\vec{R}_{\rm geod.} \cdot \langle
n,\vec{R}|\vec{\nabla}_{\vec{R}}|n,\vec{R}\rangle \ ,
\eqno(3.52)$$ so that $\gamma_n(C)$ for open cycle given by (3.47)
becomes equal to the geometric phase for the closed circuit of
(3.52). Although the above analysis has been made by using
equation (3.47) under the adiabatic approximation a
straightforward analogous treatment can be made for the
nonadiabatic case by the use of equation (3.50).

\subsubsection{The Use of Pancharatnam Phase in Neutron
Interferometry}

We discuss here a special experiment in neutron interferometry.
Using polarized neutrons Wagh et al. [60,61] have determined
phases as well as interference amplitudes for {\em noncyclic}
spinor evolution in {\em static magnetic fields}.
Interferometrically, the noncyclic phase ought to be determined
from the shift between U(1) interference without and with the
Hamiltonian affecting the evolution [60]. Wagh et al. [61] have
presented the first observation of the {\em nonclycic} phase for
neutrons and the associated amplitude of interference. We review
here the basic theory [60,37] on which these experiments are
based.

The time development of the spin-$\frac{1}{2}$ particle which
performs a precession in angle $\theta$ around a constant magnetic
field $\vec{B}=B\hat{Z}$ is given at time $t$ by the wavefunction
$$|\psi_t\rangle=e^{-i(\phi/2)}\cos\left(\frac{\theta}{2}\right)|+
Z\rangle +
e^{i(\phi/2)}\sin\left(\frac{\theta}{2}\right)|-Z\rangle
\eqno(3.53)$$ where the Larmor phase $\phi$ is given by
$$\phi=-2\mu Bt/\hbar \ , \eqno(3.54)$$ and $\mu$ is the magnetic
dipole of the particle. Here we assumed that at time $t=0$ the
wavefunction of the particle is given by
$$|\psi_0\rangle=\cos\left(\frac{\theta}{2}\right)|+Z\rangle
+\sin\left(\frac{\theta}{2}\right)|-Z\rangle \ . \eqno(3.55)$$ One
should notice that after a full rotation $(\phi=2\pi)$ the
wavefunction inverses its sign. This result is due to the nature
of spin-$\frac{1}{2}$ particle.

The Pancharatnam phase difference between $|\psi_0\rangle$ and
$|\psi_t\rangle$ is given after a straightforward calculation
[37,60] as
$$\alpha=\arg\langle\psi_0|\psi_t\rangle=-\arctan\left\{\cos\theta\tan(\phi/2)\right\}
\ . \eqno(3.56)$$

The phase $\alpha$ can be tested in neutron interferometry by
dividing the neutron beam into one beam which is exposed to the
magnetic field $\vec{B}$ while applying a variable U(1) phase
$\chi$ to the other beam. The interference intensity measured by a
detector in one outport of the interferometer can be represented
as
$$I \propto 1 + \nu \cos (\chi-\phi) \eqno(3.57)$$
where
$$\nu=|\langle
\psi_0|\psi_t\rangle|=\left[1-\sin^2\theta\sin^2(\phi/2)\right]^{\frac{1}{2}}
\eqno(3.58)$$

In the above calculation we find that the time development of the
wavefunction is a function of only one parameter. Therefore we can
use the following form for the geometric phase [37,40,60]
$$\gamma_{\rm geom.}(C)=\arg\langle
\psi_0|\psi_t\rangle-Im \int^t_0
\langle\psi_{t'}|\frac{d}{dt}|\psi_{t'}\rangle dt' \eqno(3.59)$$
where the second term on the right side of equation (3.59) is the
dynamical phase
$$\alpha_D=-\left(\frac{\phi}{2}\right)\cos\theta \eqno(3.60)$$
and the geometrical phase for the noncyclic evolution is given by
$$\gamma_{\rm geom.} = \alpha-\alpha_D \eqno(3.61)$$
The above equations are reduced to simple forms for
$\theta=0,\pi$. The we get
$$\nu=1; \ \alpha_D=\pm(\phi/2); \ \alpha = \pm(\phi/2)
\eqno(3.62)$$ and for this case the geometric phase vanishes. For
$\theta=\pi/2$ we find that $\langle\psi_0|\psi_t\rangle$ is real
and the Pancharatnam phase is undefined. Excluding these special
cases, we find that for the general case both dynamical and
geometrical phases are obtained by the use of equations
(3.60-3.61).

\section*{Chapter 4. \ Measurements of Topological Phases}
\addtocounter{section}{1} \addtocounter{subsection}{-4}
\subsection{Topological Phases in Molecular Jahn-Teller System}

General treatments of Topological phases in molecular spectroscopy
are quite complicated [2-3,94-103] and needs expertise of this
scientific area. We demonstrate basic concepts of molecular
topological phases by giving here a simple analysis of Jahn-Teller
effect [101-102].

We treat here a doubly degenerate molecular electronic states
$|\psi_1\rangle$ and $|\psi_2\rangle$ where one needs to consider
a combination of $|\psi_1\rangle$ and $|\psi_2\rangle$ for their
interaction with the molecular vibrations. Such interactions are
beyond the Born-Oppenheimer approximation [101-102]. In
Jahn-Teller system one considers, for example, triatomic molecules
(ions) in which there is a coupling between the degenerate
electronic states and the nuclear vibrations, or a group of atoms
in crystals. For simplicity of the present treatment we consider a
coupling between the degenerate electronic state and two normal
coordinate vibrations $Q_a$ and $Q_b$, which have the same
frequency $\omega$. We consider interaction $V_{in}$ which is
linear in the $Q_a$ and $Q_b$ coordinates and in the form
$$V_{int} = - K\left(\begin{array}{cc} -Q_a & Q_b \\ Q_b &
Q_a\end{array}\right) \ , \eqno(4.1)$$ as required by the
Jahn-Teller analysis. The constant $K$ represents the strength of
the coupling between the degenerate electronic states and the two
normal vibration coordinates. The effective Hamiltonian of the
normal vibration is given by
$$H_{\rm eff}=\frac{P_a^2 + P_b^2}{2\mu} +
\frac{\mu\omega^2}{2}(Q_a^2+Q_b^2)-K\left(\begin{array}{cc} -Q_a &
Q_b \\ Q_b & Q_a\end{array}\right) \eqno(4.2)$$ where $P_a$ and
$P_b$ are the conjugate momenta to $Q_a$ and $Q_b$, respectively.
$\mu$ is the effective mass of the normal vibration. The normal
coordinates can be written
$$Q_a=\rho\cos\theta \ ; \ Q_b=\rho\sin\theta \eqno(4.3)$$
where $\rho$ is a constant and $\theta$ is variable. The last term
of equation (4.2) removes the degeneracy of the electronic
energies $\epsilon_{\pm}$ which could be considered as effective
vibrational potentials to be added to the vibrational potential of
the double degenerate normal coordinate $(Q_a,Q_b)$:
$$V_{ab}^{\pm} = \left(\frac{\mu\omega^2}{2}\right) \rho^2 +
\epsilon^\pm \eqno(4.4)$$

The positive potential $V_{ab}^+$ and the negative one $V^-_{ab}$
describe potential surfaces as a function of the coordinates
$Q_a,Q_b$ where near the degeneracy point these adiabatic surfaces
are conical.

In order to find the eigenenergies $\epsilon_\pm$ and the
electronic eigenkets $|+(\theta)\rangle$ and $|-(\theta)\rangle$,
corresponding to these eigenenergies, we solve the equation
$$-K\rho\left(\begin{array}{cc} -\cos\theta &
\sin\theta \\ \sin\theta & \cos\theta\end{array}\right)
\left(\begin{array}{c} C_1 \\ C_2 \end{array}\right) = \epsilon
\left(\begin{array}{c} C_1 \\ C_2 \end{array}\right) \ .
\eqno(4.5)
$$
The electronic energies are given as
$$\epsilon_\pm = \pm K\rho \ , \eqno(4.6)$$
and the calculation of the amplitudes $C_1,C_2$ give the
corresponding eigenkets:
$$\begin{array}{l} |+(\theta)\rangle=\cos(\frac{\theta}{2})|\psi_1\rangle -
\sin(\frac{\theta}{2}) | \psi_2\rangle , \\
|-(\theta)\rangle=\sin(\frac{\theta}{2})|\psi_1\rangle +
\cos(\frac{\theta}{2}) | \psi_2\rangle \end{array} \ .
\eqno(4.7)$$ Here we notice that electron kets change sign for
rotation by $2\pi$ around the symmetry axis
$$\left| \pm (\theta=2\pi)\rangle = (-1)\right|\pm \theta(0)\rangle
\ . \eqno(4.8)$$ Formally, this is the same as the rotation of a
$\frac{1}{2}$-spin system discussed in Section (2.1), since the
sinusoidal dependence of the kets on the rotation is given by
one-half of the rotation angle. One should notice, however, that
it is the {\em orbital state} ket of the electron system that has
changed sign and thus has nothing to do with spins. The factor -1
on the right side of equation (4.8) is shown to be related to
Berry's phase factor. Since the eigenkets of equation (4.7) are
not single valued one should find a phase factor that will make
the kets single valued. For $|-(\theta)\rangle$, for example,
define
$$| X_-\rangle = |-(\theta)\rangle e^{i\theta/2} \ . \eqno(4.9)$$
The transformation (4.9) restores the single valuedness, but now
there is a nonzero vector potential
$$A_\theta=(\frac{1}{i})\langle
X_-\left|\frac{d}{d\theta}\right|X_-\rangle=\frac{1}{2} \ .
\eqno(4.10)$$ The integration of the vector potential around a
closed circuit which includes in the circuit the origin of the
$Q_a,Q_b$ plane (which is the degeneracy point of the electron
system) gives Berry's phase factor $\pi$ which change the sign of
the wavefunction $|X_-\rangle$. It was noticed by molecular
spectroscopists that restoration of the single valuedness of the
electronic eigenstate leads to nonzero vector potentials
[2,94-97].

The dynamical motion of the normal coordinates $Q_a, Q_b$ (given
by $\rho,\theta$), under the potentials of equation (4.4), are
obtained by solving the two-dimensional Schrodinger equation
$$\left\{ -
\left(\frac{\hbar^2}{2\mu}\right)\left(\frac{\partial^2}{\partial\rho^2}
+ \left(\frac{1}{\rho}\right)\frac{\partial}{\partial \rho} +
\left(\frac{1}{\rho^2}\right)
\frac{\partial^2}{\partial\theta^2}\right) +
V^\pm_{ab}(\rho)\right\}\psi(\rho,\theta) = E\psi(\rho,\theta)
\eqno(4.11)$$
$$\psi(\rho,\theta)=\left\{\begin{array}{l}
\phi_+(\rho,\theta)|+(\theta)\rangle \\
\phi_-(\rho,\theta)|-(\theta)\rangle \end{array} \right.
\eqno(4.12)$$ Since equation (4.11) is a differential equation
with separable variables we can solve it by choosing the nucleus
wavefunction as
$$\phi_\pm(\rho,\theta)=f_\pm(\rho)e^{ij\theta} \ . \eqno(4.13)$$

The state ket of the electron system changes sign when $\theta$ is
rotated by $2\pi$. Therefore in order to cancel this sign change,
$\phi_\pm(\rho,\theta)$ must change its sign to preserve the
single valued nature of $\psi(\rho,\theta)$. From this requirement
we conclude that $\exp(ij2\pi)=-1$. That is
$$ j= \ {\rm half \ integer}, \ (\ell = {\rm integer}) \ .
\eqno(4.14)$$ The presence of the Berry's has changed the quantum
number that characterized the rotation of atoms, from the usual
integer number to half integer ones!

The low energy excitations in the potential as given by
$$E_{nj} \simeq (n+\frac{1}{2})\hbar\omega + \frac{\hbar^2(j^2+
\frac{1}{4})}{2\mu\rho_0^2} \ ; \ n \ {\rm integer}; \ j= {\rm
half \ integer} \ . \eqno(4.15)$$ The result is given by the sum
of the energies of vibration along the radial direction and those
of rotation in the plane of the normal mode coordinates.

We should notice that the demand for single valuedness in the
above treatment has been made for the total molecular wavefunction
describing a {\em closed} system, i.e., a system for which the
total Hamiltonian is conserved. In the treatments of Berry's phase
given in Sections (3.1)-(3.3) we have assumed an Hamiltonian which
is a function of time. For such {\em open} systems we have not the
requirement of single valuedness which will be similar to that of
molecular spectroscopy.

\subsection{Measurements of Topological Phases with Polarized Light}
\subsubsection{Berry's Phases on Poincare Sphere}

Let us give a simple explanation for the description of a Poincare
sphere by Stokes parameters [104] which are equivalent to the
Bloch vector. The em field of a monochromatic plane polarized
light propagating in the $z$ direction can be given as
$$E_x = a_1 \cos (\tau +\delta_1) \ ; \ E_y=a_2\cos(\tau+\delta_2)
\eqno(4.16)$$ where $a_1$ and $a_2$ are the amplitudes in the $x$
and $y$ directions. $\tau$ is the variable phase factor while
$\delta_1$ and $\delta_2$ are constants. By using a simple algebra
[104] one obtains the equation of an ellipse
$$\left(\frac{E_x}{a_1}\right)^2 + \left(\frac{E_y}{a_2}\right)^2
- 2 \left(\frac{E_x}{a_1}\right)\left(\frac{E_y}{a_2}\right)
\cos\delta = \sin^2\delta \eqno(4.17)$$ where
$$\delta = \delta_1-\delta_2 \ . \eqno(4.18)$$

The ellipse will be reduced to {\em linearly polarized light} when
$\delta=\delta_2-\delta_1 = m\pi (m=0,\pm1,\pm2 \dots)$. The
ellipse is reduced to right-handed (left-handed) {\em circularly
polarized} wave for $\delta=2m\pi+\pi/2(2m\pi-\pi/2)$ [104].
Instead of using the real form of equation (4.16) it is more
convenient to describe the em field as a complex Jones vector
[105]
$$\begin{array}{l}E_x({\rm complex}) =
a_1e^{i(\tau+\delta_1)}=A_1\\[0.25cm]
E_y({\rm complex}) = a_2e^{i(\tau+\delta_2)}=A_2
\end{array} \ . \eqno(4.19)$$
Jones vector is described as a complex spinor vector
$\left(\begin{array}{c} A_1 \\ A_2 \end{array}\right)$. The Stokes
parameters of a plane monochromatic wave are described by the four
quantities
$$\begin{array}{l} S=a_1^2 + a^2_2 = |A_1|^2 + |A_2|^2, \\
S_1=a^2_1 - a^2_2 = |A_1|^2 - |A_2|^2, \\
S_2 = 2a_1a_2\cos(\delta) = A_1^*A_2 + A_1A_2*, \\
S_3=2a_1a_2 \sin\delta =(A_1A_2^*-A^*_1A_2)/i \ .
\end{array} \eqno(4.20)$$
The parameter $S$ is proportional to the intensity of the wave and
this parameter can be normalized to 1 for unitary transformations.
Then the Stokes parameters are equivalent {\em mathematically} to
the Bloch vector components described in Section (2.1.3). For
nonunitary transformations the magntiude of the Bloch vector, or
equivalently the radius of the Poincare sphere, is decreasing then
the parameters become important. However, by using the
Pancharatnam phase approach the topological phase obtained for
nonunitary transformations are similar to those of unitary
transformations. Using Stokes parameters right-handed
(left-handed) circular polarization is presented by the north
(south) pole of the Poincare sphere. Linear polarization is
represented by points in the equatorial plane [105].

Jones vectors for right and left polarized light waves are given
by [105]:
$$\hat{R}=\frac{1}{\sqrt{2}}
\left(\begin{array}{l}1\\-i\end{array}\right) \ ; \ \hat{L}=
\frac{1}{\sqrt{2}}\left(\begin{array}{l}1\\i\end{array}\right).
\eqno(4.21)$$

These two circular polarizations are mutually orthogonal in the
sense $\hat{R}^\dagger\hat{L}=0$. Any pair of orthogonal Jones
vectors can be used as a basis of the mathematic space spanned by
the Jones vector. This description corresponds to spinors of
two-level system where the two levels are here two orthogonal
complex polarization states. Elliptic polarization can be
represented as the following Jones vector
$$\hat{J}=\left(\begin{array}{ll} \cos & \psi\\e^{i\delta}
& \sin \psi\end{array}\right) \ . \eqno(4.22)$$ By operating on
the polarization states with different optical apparatus (e.g.,
retardation waveplates, active rotating materials, polarizers,
etc.) the two components of the Jones vector are transformed
differently [105].

Bhandari and Samuel [43] observed Poincare topological phases by
moving the polarization state of a monochromatic field on the
Poincare sphere. For this purpose they have used quarter phase
plates (which produces a difference in phase which is a quarter of
$2\pi$ between two orthogonal polarizations) and optical active
medium (which causes a certain rotation of the plane of polarized
light). They also exchanged the optical active material by
polaroids and in this wave have shown that the Pancharatnam
geometric phase can be obtained also for nonunitary
transformations. In either case the beam comes back with an added
topological phase which is equal to the {\em half the solid angle}
substended by the area of the circuit on the Poincare sphere. In
principal due to a large dynamical phase it is very difficult to
measure the topological phases. Therefore the real measurements
[43] were obtained as the {\em difference in the topological
phases} for two circuits which have the same dynamical phase.
There are various articles reporting observations of polarization
states on the Poincare sphere [43-45, 106].

\subsubsection{Topological Phases of Photons in Momentum Space}

There is an analogy between the change in direction $(\hat{k}_x,
\hat{k}_y, \hat{k}_z$) of the photon by using externally slowly
varying parameters and by adiabatic change in the direction of the
magnetic field $(B_x, B_y, B_z)$ where the geometrical phase for
the latter case has been treated by Berry, as presented in Section
(3.1). If there is nothing to change the sign of the helicity of
the photon the helicity quantum number is an adiabatic invariant.

When the photon propagates smoothly down a helical waveguide (or
more practically through a fibre), $\vec{k}$ is constrained to
remain parallel to the local axis of the waveguide, since the
momentum of the photon is in this direction. The geometry of a
helical path of a waveguide (or fibre with a unity winding number)
constrains $\vec{k}$ to trace out a loop on the surface of a
sphere in parameter space $(k_x, k_y, k_z)$ where the origin
$\vec{k}=0$ of this space is singular [compare $\vec{k}=0$ with
the singular point $\vec{B}=0$ for the case treated by Berry [4]].
Following Berry's argument [as presented in Section (3.1)] we
obtain here a geometrical phase
$$\gamma(C)=-\sigma\Omega(C) \eqno(4.23)$$
where $\sigma=\pm 1$ is the helicity number and $\Omega(C)$ is the
solid angle substended by the loop $C$ with respect to
$\vec{k}=0$. For a circle one gets [4]:
$$\Omega(C) = 2\pi N(1-\cos\theta) \eqno(4.24)$$
where $N$ is the winding number of the helix and $\theta$ is the
angle between the local fibre axis and the axis of the helix (the
`pitch angle') of the helix.

In the experiment [35] a linearly polarized laser has been focused
into a single-mode isotropic fibre which is wound into the shape
of a helix. The fibre output points are at the same direction as
the input. The output light is analyzed to determine its axis of
linear polarizer relative to that of the input light.

Let the initial state be represented by
$$| X \rangle = 2^{-\frac{1}{2}} (|+ \rangle + |-\rangle)
\eqno(4.25)$$ where $|\pm\rangle$ are the eigenstates of the
helicity $\sigma=\pm 1$. After propagating through the helix, the
final state at the output of the fibre, if we ignore dynamical and
birefrienges effects, is
$$| X' \rangle =2^{-\frac{1}{2}} (\exp(i\gamma_+)|+\rangle +
\exp(-i\gamma_+)|-\rangle) \eqno(4.26)$$ Herre $\gamma_+$ is the
phase for $\sigma=+1$. Therefore $|\langle X|X'\rangle|^2 =
\cos^2\gamma_+$. By Malus law, this implies that the plane
polarization has been rotated by an angle which is equal to
$\gamma_+$. Chiao and Wu [34] have discussed the validity
conditions for the adiabatic approximation and have considered
also other cases of polarized light by which Berry's geometric
phase can be measured. The elimination of dynamical and other
optical activities in the experiments have been discussed [34-35]
including the role of geometric phase as a quantum or classical
effect [33-35, 107-109].

The topological phases of photons in momentum space have been
realized by various constructions of interferometers: 1)
Non-planar Mach-Zhender interferometers where the geometric phase
is observed as fringe shift [110-111] In such interferometers
additive effects of Poincare sphere and helicity momentum phase
have been observed [112]. 2) Interferometers with fibre loops have
been built into ring interferometer [56] and in Sagnac
interferometer [57] and have been used by coincidence counters to
determine the geometric phase for a single photon and of a photon
pair from parametric down conversion [113]. 3) The effects of
reflections on quantum phases have been discussed [114].

\subsection{Neutron Interferometry}

Berry's original idea to observe the geometric phase for a spin
particle around the magnetic field {\em under the adiabatic
approximation} has been realized experimentally by Bitter and
Dubbers [46]. They have used polarized cold electrons ($\lambda =
8 \AA)$ propagating along a helical magnetic field which allows
for an easy fulfillment of the adiabaticity conditions.

In the usual interferometry experiments with neutrons the
adiabatic approximation is not valid [58,115]. There are various
articles treating the phase shifts of spinor wavefunctions
interacting with magnetic fields [116-122]. Phases in coupled
neutron interference loops and off-diagonal geometric phases in
polarized neutron interferometry have been observed [123-124].

An analysis of topological phases in neutron interferometry has
been discussed in Section  3.4.3, where such analysis with the
corresponding measurements have been related to Pancharatnam phase
analysis, which is valid also for noncyclic geometric phase.

\subsection{Observations of Lorentz-Group Topological Phases}

Kitano and Yabuzaki (KY) [51] have described Lorentz-Group Berry's
phases observations in polarization optics. They have used in
their analysis nonunitary Lorentz-Group transformations which are
different from the unitary transformations described by Chiao and
Jordan (CJ) [50] for squeezed $em$ fields [125] as those described
in Section 2.1.6. The Lorentz-Group generators used by $KY$ are
given by the matrices:
$$K_1 = i\sigma_z/2 = \left(\begin{array}{ll} \frac{i}{2} & 0\\0 &
-\frac{i}{2}\end{array}\right) \ ; K_2 = i\sigma_x/2 =
\left(\begin{array}{ll}   0 & \frac{i}{2} \\
\frac{i}{2} & 0 \end{array}\right) \ ;K_3 = J_3 = \sigma_y/2 =
\left(\begin{array}{ll}   0 & -\frac{i}{2} \\
\frac{i}{2} & 0 \end{array}\right) \ . \eqno(4.27)$$ The CR of
these operators are the same like those of (2.9), but since here
the generators $K_1$ and $K_2$ are non-hermitian the present
classical polarization optics system is basically different from
that of the quantized squeezed $em$ fields.

KY have used the transformation
$$T=S_4S_3S_2S_1 = \exp(irK_2)\exp(isK_1)\exp(-isK_2)\exp(-irK_1)
\eqno(4.28)$$ This transformation is the same as that suggested by
CJ but here the operators $S_1, S_2, S_3$ and $S_4$ are not
unitary. A straightforward calculation shows that
$$\exp(isK_1) =  \left(\begin{array}{ll}   \exp(-\frac{s}{2})& 0  \\
0 & \exp(\frac{s}{2}) \end{array}\right)\ ; \ \exp(irK_2) =
=\left(\begin{array}{ll}   \cosh(\frac{r}{2})& -\sinh(\frac{r}{2})  \\
-\sinh(\frac{r}{2})  & \cosh(\frac{r}{2}) \end{array}\right) \ .
\eqno(4.29)$$ In order to realize the transformation given by
equation (4.28) KY have squeezed the space of polarization states
of light by linear polarizers. A linearly polarized light wave
propagating along the $z$ direction can be described by a pair of
real components $E_x$ and $E_y$ arranged as a column vector. Then
the effect of a linear polarizer whose optical axes are set
parallel to the coordinate axes can be represented by the $2\times
2$ matrix
$$S = {\rm diag} (t_x, t_y)=t_a \left(\begin{array}{ll}  \kappa & 0 \\
0  & \kappa^{-1} \end{array}\right)\eqno(4.30) $$ where $t_x(t_y)$
is the transmission coefficient for the $x(y)$ component of the
field. We note that $S$ is composed of two parts: the isotropic
loss $t_a=(t_xt_y)^{\frac{1}{2}}$. Since KY were not interested in
the absolute intensities but in the polarization states of light,
they neglected the isotropic losses.

The realization of the $T$ transformation has been obtained [51]
by using a sequence of polarization squeezing. If the parameters
are chosen so that $\tanh(r)=\sinh(s)$, then $T$ leads to a cyclic
transformation
$$T\propto \exp(-i\phi J_3) \eqno(4.31)$$
where the rotation angle $\phi/2$ is given by the relation
$$\sin\phi=\tanh^2(r)\eqno(4.32)$$
It has been shown that when the extinction ratio of each polarizer
is adjusted properly the system becomes equivalent to a (lossy)
polarization rotator.

The basic idea of Lorentz-Group Berry's topological phase is that
such phase will be  given by
$$\theta_B=\oint d\Omega \eqno(4.33)$$
where the integral is over the area of the hyperboloid (Lorentz
metric) enclosed by the loop [82]. Although the measurements of
the phase shift $\phi$ measured by KY are in good agreement with
their theoretical calculations, the phase $\phi$ measured by them
includes the summation of geometrical and dynamical phases.

Quantum systems exhibiting SU(2) or SU(1,1) dynamical symmetry
including theoretical calculations of geometrical and dynamical
phases have been analyzed by Cervero and Lejarreta [126-127].
Experimental observation of Berry's phases of the Lorenz group has
been reported by Swensmark and Dimon [128]. They have analyzed a
nonlinear system, which when tuned in the vicinity of an
instability shows a response similar to a Lorentz group. They have
used a number of successive transformations combined in such a way
that they make a closed loop on the hyperboloid. Their
measurements of the Berry's phase are in a good agreement with the
use of equation (4.33).

\section*{Chapter 5. Summary and Discussion}
\addtocounter{section}{1} \addtocounter{subsection}{-5}

Geometric phases are intrinsically related to fibre-bundle
topological theories. Usually topology is considered as the
mathematical basis for general relativity and for certain fields
in high energy physics. However, deep understanding of geometric
phases related to time development by Schrodinger equation and
interference effects in optics needs also the use of these
analytical methods. Following this idea basic elements of topology
are explained in Chapter 2, including the description of various
manifolds, Cartan exterior algebra, fibre-bundles theories and
differential geometry basic concepts. The discussions presented in
this chapter should be helpful in analyzing complicated
topological phases, e.g., those related to quantum computation.
Also we demonstrate close connections between topological phases
and topological singularities.

In Chapter 3 we describe the fundamental analysis made by Berry
for topological phases related to `magnate monopoles'.
Generalization of Berry's phases to nonadiabatic time development
and their relation to fibre-bundle theories are discussed. The
analysis of Berry's phase for $n$-level atomic system is treated
by the use of Kahler metric. The relation between rays in quantum
mechanics and Berry's phase is explained by using fibre-bundle
theory. The use of Pancharatnam phases for noncyclic quantum
evolution and for optical systems in relation to geometrical phase
is discussed. The problem of gauge invariance in relation to
Pancharatnam phase is emphasized, and solutions for various
systems are given.

In Chapter 4 various measurements of geometrical phases are
described. The relation between deviation from Born-Oppenheimer
approximations nd geometric phases in molecular spectroscopy is
emphasized. A simple analysis is described for a special system
known as the Jahn-Teller effect. Although geometric phases
appeared in molecular studies quite long ago they are basically
different from those described by Berry.

The topological phases known as Berry's phases are for open
systems (with time dependent Hamiltonian) while the molecular
system of nuclei plus electrons is closed [total Hamiltonian is
conserved]. Measurements of geometric phases in polarization
optics related to SU(2) group are described. The appearance of
geometric phases in many experiments performed in this field are
analysed both for closed circuits following Berry's approach and
for nonclosed circuits following Pancharatnam formalism. Basic
experiments in neutron interferometry are discussed. A short
analysis of geometric phases which are related to the Lorentz
group is given.

It is not possible in one Review to cover all the interesting
theories and experiments which have been developed for analysing
geometrical phases phenomena. We have described here only some
selected topics in this field. There are many other theoretical
and experimental [129-186] articles in which different topics
and/or different analysis of topological phases have been
described. \pagebreak

\noindent{\bf References}
\begin{enumerate}
\item Aharonov Y and Bohm D 1959 {\em Phys.\ Rev.} {\bf 115} 485
\item Herzberg G. and Longuet-Higgins H C 1963 {\em Discussion
Farady Soc.} {\bf 35} 77
\item Mead C A and Truhlar D G 1979 {\em J. Chem\ Phys.} {\bf 70}
2284
\item Berry M V 1984 {\em Proc.\ R.\ Soc.} A {\bf 392} 45
\item Eguchi T, Gilkey P B and Hanson A J 1980 {\em Physics
Reports} {\bf 66} 213
\item Frankel T 2001 {\em The Geometry of Physics} (Cambridge:
Cambridge University Press)
\item Chern S S, Chen W H and Lam K S 1998 {\em Lectures on
Differential Geometry} (Singapore: World Scientific)
\item Nakahara M 1989 {\em Geometry, Topology and Physics}
(Bristol: Institute of Physics Publishing)
\item Nash C and Sen S 1983 {\em Topology and Geometry for
Physicists} (London: Academic Press)
\item Simon B 1983 {\em Phys.\ Rev.\ Lett.} {\bf 51} 2167
\item Aharonov Y and Anandan J 1987 {\em Phys.\ Rev.\ Lett.} {\bf
58} 1593
\item Wilczek F and Zee A 1984 {\em Phys.\ Rev.\ Lett.} {\bf 52}
2111
\item Yang C N and Mills R L 1954 {\em Phys. Rev.} {\bf 96} 191 
\item Hong-Mo C and Tsun T S 1993 {\em Some Elementary Gauge Theory
Concepts} (Singapore: World Scientific)
\item Jackiw R 1988 {\em Int. J. Mod.\ Phys.} A {\bf 3} 285
\item Mostafazadeh A 1997 {\em Phys. Rev.} A {\bf 55} 1653
\item Samuel J and Dhar A 2001 {\em Phys.\ Rev.\ Lett.} {\bf 87}
260401
\item Gubarev F V and Zakharov V I 2002 {\em Int.\ J.\ Mod.\
Phys.} {\bf 17} 157
\item Pistolesi F and Manini N 2000 {\em Phys.\ Rev.\ Lett.} {\bf
85} 1585
\item Lauber H-M, Weidenhammer P and Dubbers D 1994 {\em Phys.\
Rev.\ Lett.} {\bf 72} 1004
\item Manolopoulos D E and Child M S 1999 {\em Phys. Rev. Lett.}
{\bf 82} 2223
\item Pachos J, Zanardi P and Rasetti M 1999 {\em Phys. Rev.\
A} {\bf 61} 010305
\item Pachos J and Chountasis S 2000 {\em Phys. Rev.} A {\bf 62}
052318
\item Pachos J and Zanardy P 2001 {\em J. Mod.\ Phys.} B {\bf 15}
1257 
\item Margolin A E, Strazhev V I and Tregubovich A Y 2002 {\em
Phys. Lett} A {\bf 303} 131
\item Margolin A E, Strazhev V I and Tregubovich A Y 2003 {\em
Phys. Lett.} A {\bf 312} 296
\item Hannay J H 1985 {\em J. Phys. A: Math. Gen.} {\bf 18} 221.
\item Berry M V 1985 {\em J. Phys. A: Math. Gen.} {\bf 18} 15.
\item Pati A K 1998 {\em Annals of Physics} {\bf 270} 178.
\item Cerver\'o J M and Lejarreta D 1990 {\em Quantum Opt.} {\bf 2}
339
\item Anandan J and Aharonov Y 1988 {\em Phys.\ Rev.\ D} {\bf 38}
1863
\item Anandan J and Stodolsky L 1987 {\em Phys.\ Rev.\ D} {\bf 35}
2597
\item Jordan T F 1987 {\em J. Math. Phys.} {\bf 28} 1759
\item Chiao R Y and Wu Y  1986 {\em Phys.\ Rev.\ Lett} {\bf 57}
933
\item Tomita A and Chiao R Y 1986 {\em Phys. Rev.\ Lett.} {\bf 57} 937
\item Pancharatnam S 1956 {\em Proc.\ Ind.\ Acad.\ Sci.} A {\bf
44} 247
\item Sjoqvist E 2001 {\em Phys.\ Rev.\ A} {\bf 63} 035602
\item Jordan T F 1988 {\em Phys. Rev. A} {\bf 38} 1590
\item Samuel J and Bhandari R 1988 {\em Phys.\ Rev.\ Lett.} {\bf
60} 2339
\item Mukunda N and Simon R 1993 {\em Annals of Physics} {\bf 228}
205 
\item Aitchison I J R and Wanelik K 1992 {\em Proc.\ R.\ Soc.\
Lond.} {\bf 439} 25
\item Pati A K 1995 {\em Phys.\ Rev.\ A} {\bf 52} 2576
\item Bhandari R and Samuel J 1988 {\em Phys.\ Rev.\ Lett.} {\bf
60} 1211
\item Bhandari R 1988 {\em Phys.\ Lett.\ A} {\bf 133} 1; 1990 {\em
Phys.\ Lett.} A {\bf 143} 170; 1997 {\em Phys.\ Lett.} A {\bf 157}
221; 1992 {\em Phys.\ Lett.} A {\bf 171} 262; 1993 {\em Phys.
Lett.} A {\bf 180} 15
\item Bhandari R 1997 {\em Physics Reports} {\bf 281} 1.
\item Bitter T and Dubbers D 1987 {\em Phys.\ Rev.\ Lett.} {\bf
59} 251
\item Suter D, Mueller K T and Pines A 1988 {\em Phys.\ Rev.\
Lett.} {\bf 60} 1218
\item Simon R, Kimble H J and Sudarshan E C G 1988 {\em Phys.\
Rev.\ Lett.} {\bf 61} 19
\item Agarwal G S 1991 {\em Optics Commun.} {\bf 82} 213.
\item Chiao R Y and Jordan T F 1988 {\em Phys.\ Lett.} A {\bf 132}
77
\item Kitano M and Yabuzaki T 1989 {\em Phys. Lett. A} {\bf 142}
321 
\item Hariharan P, Roy M, Robinson P A and O'Byrne J W  1993 {\em
J. Mod. Optics} {\bf 40} 871
\item Kwiat P G and Chiao R Y 1991 {\em Phys.\ Rev.\ Lett.} {\bf
66} 588
\item Brendel J, Dultz W and Martienssen W 1995 {\em Phys.\ Rev.}
A {\bf 52} 2551
\item Grayson T P, Torgerson J R and Barbosa G A 1994 {\em Phys.\
Rev} A {\bf 49} 626
\item Frins E M and Dultz W 1997 {\em Optics Commun,} {\bf 136}
354
\item Senthilkumaran P, Culshaw B and Thursby G 2000 {\em J.\ Opt.\
soc. Am.\ B} {\bf 17} 1914
\item Rauch H and Werner S A 2000 {\em Nuetron Interferometry}
(Oxford: Clarendon Press)
\item Wagh A G and Rakhecha V C 1990 {\em Phys.\ Lett.\ A} {\bf
146} 369; 1992 {\em Phys.\ Lett. A} {\bf 170} 71; 1995 {\em Phys.\
Lett.} A {\bf 197} 107
\item Wagh A G, Rakhecha V C, Fischer P and Joffe A 1998 {\em
Phys.\ Rev.\ Lett.} {\bf 81} 1992 
\item Wagh A G, Rakhecha V C, Summhammer J, Badurek G, Weinfurter
H, Allman B E, Kaiser H, Hamacher K, Jacobson D L and Werner S A
1997 {\em Phys.\ Rev.\ Lett.} {\bf 78} 755
\item Sanders B C, De Guise H, Bartlett S D and Zhang W 2001 {\em
Phys.\ Rev.\ Lett.} {\bf 86} 369
\item Strahov E 2001 {\em J. Math. Phys.} {\bf 34} 2008
\item Ben-Aryeh Y 2002 {\em J.\ Mod.\ Opt.} {\bf 49} 207
\item Ben-Aryeh Y 2003 {\em Optics and Spectroscopy} {\bf 94} 783
\item Hessmo B and Sj\"{o}qvist E 2000 {\em Phys.\ Rev.} A {\bf 62}
062301
\item Milman P and Mosseri R 2003 {\em Phys. Rev.\ Lett.} {\bf 90}
230403
\item Sj\"oqvist E 2000 {\em Phys.\ Rev. A} {\bf 62} 022109
\item Ericsson M, Sj\"ovist E, Br\"annlund J, Oi D K L and Pati A
K 2003 {\em Phys.\ Rev.} A {\bf 67} 020101 
\item Filipp S and Sjoqvist E 2003 {\em Phys.\ Rev.\ Lett.} {\bf
90} 050403
\item Nozir A, Spiller T P and Munro W J 2002 {\em Phys.\ Rev.\ A}
{\bf 65} 042303
\item Friedenauer J and Sj\"oqvist E 2003 {\em Phys.\ Rev.\ A}
{\bf 67} 024303
\item Jones J A, Vedral V, Ekert A and Castagnoll G 2000 {\em
Nature} {\bf 403} 869
\item Falci G, Fazio R, Palma G M, Siewert J and Vedral V 2000 {\em
Nature} {\bf 407} 355
\item Shapere A and Wilczek F 1989 {\em Geometric Phases in
Physics} (Singapore: World Scientific)
\item Yoshioka D 2002 {\em The Quantum Hall Effect} (Berlin:
Springer)
\item Anandan J and Aharonov Y 1990 {\em Phys.\ Rev.\ Lett.} {\bf
65} 1697
\item Provost J P and Vallee G 1980 {\em Commun.\ Math.\ Phys.}
{\bf 76} 289
\item Pati A K and Joshi A 1993 {\em Europhys.\ Lett.} {\bf 21}
723
\item Abe S 1993 {\em Phys.\ Rev.\ A} {\bf 48} 4102
\item Berry M V 1989 (page 7, in Ref. 75) 
\item Dattoli G, Dipace A and Torre A 1986 {\em Phys.\ Rev. A}
{\bf 33} 4387
\item Bohm A, Boya L J and Kendrik B 1991 {\em Phys.\ Rev.\ A}
{\bf 43} 1206
\item Dirac A M 1931 {\em Proc.\ Roy.\ Soc.\ A} {\bf 133} 60
\item Jackson J D 1975 {\em Classical Electrodynamics} (New-York:
John Wiley)
\item Goddard P and Olive D 1978 {\em Rep.\ Prog.\ Phys.} {\bf 41}
91
\item Wu T and Yang C N 1976 {\em Nucl. Phys. B} {\bf 107} 365
\item Jackiw R 1980 {\em Annals of Physics} {\bf 129} 183
\item Cirelli R and Lanzavecchia P 1984 {\em Nuvo Cimento} {\bf
79} 271
\item Cirelli R, Mania A and Pizzocchero L 1991 {\em Int.
J. Mod. Phys.} {\bf 6} 2133
\item Flaherty E J 1976 {\em Hermitian and K\"ahlerian Geometry in
Relativity} (Berlin: Springer in Lecture Notes in Physics)
\item Page D N 1987 {\em Phys.\ Rev.} A {\bf 36} 3479
\item de Bolaviega G G and Sj\"oqvist A 1998 {\em Am. J.\ Phys.}
{\bf 66} 431
\item Mead C A 1992 {\em Rev.\ Mod.\ Phys.} {\bf 64} 51
\item Longuet-Higgins H C 1975 {\em Proc.\ R.\ Soc.\ Lond.} A
{\bf 344} 147
\item Mead C A 1979 {\em J.\ Chem.\ Phys.} {\bf 70} 2276
\item Mead C A 1987 {\em Phys.\ Rev.\ Lett.} {\bf 59} 161
\item Sjoqvist E and Hedstr\"om 1997 {\em Phys.\ Rev.\ A} {\bf 56}
3417
\item Sjoqvist E 2002 {\em Phys.\ Rev. Lett.} {\bf 89} 210401
\item Avron J E 2000 {\em Phys.\ Rev.} A {\bf 62} 062504
\item Aitchison I J R 1988 {\em Physica Scripta} {\bf T23} 12
\item Sakurai J J {\em Modern Quantum Mechanics} 1994 (New York:
Addison-Wesley)
\item Kuratsuji H and Ida S 1985 {\em Prog.\ Theor.\ Phys.} {\bf
74} 439
\item Born M and Wolf E 1965 {\em Principles of Optics} (Oxford:
Pergamon Press)
\item Yariv A 1991 {\em Optical Electronics} (New York: Saunders
College Publishing)
\item Chyba T H, Wang L J, Mandel L and Simon R 1988 {\em Optics
Letters} {\bf 13} 562
\item Berry M V 1987 {\em Nature} {\bf 326} 277
\item Lipson S G 1990 {\em Optics Letters} {\bf 15} 154 
\item Ross J N 1984 {\em Optical and Quantum electronics} {\bf 16}
455
\item Shen J-Q and  Ma L-H 2003 {\em Phys.\ Lett.\ A} {\bf 308}
355
\item Chiao R Y, Antaramian A, Ganga K M, Jiao H, Wilkinson S R
and Nathel H 1988 {\em Phys.\ Rev.\ A} {\bf 60} 1214
\item Jiao H, Wilkinson S R, Chiao R Y and Nathel H 1989 {\em
Phys.\ Rev.} A {\bf 39} 3475
\item Siebert K J, Schmitzer H and Dultz W 2002 {\em Phys.\ Lett.}
A {\bf 300} 341
\item Jordan T F 1988 {\em Phys.\ Rev.\ Lett.} {\bf 60} 1584.
\item Klein A G and Werner S A 1983 {\em Rep. Prog. Phys.} {\bf
46} 259
\item Rauch H, Zeilinger A, Badurek G, Wilfing A, Bauspiess W and
Bonse U 1975 {\em Phys. Lett.} {\bf 54} 425
\item Klein A G and Opat G I 1976 {\em Phys.\ Rev.\ Lett.} {\bf
37} 238
\item Mezei F 1988 {\em Physica} B {\bf 151} 74
\item Badurek G, Rauch H, Zeilinger A, Bauspiess W and Bonse A 1976
{\em Phys.\ Rev.} D {\bf 14} 1177 
\item Summhammer J, Budarek G, Rauch H, Kischko U and Zeilinger A
1983 {\em Phys.\ Rev.\ A} {\bf 27} 2523
\item Werner S A, Colella R, Overhauser A W and Eagen C F 1975
{\em Phys.\ Rev.\ Lett.} {\bf 35} 1053
\item Aharonov Y and Susskind L 1967 {\em Phys.\ Rev.} {\bf 158}
1237
\item Hasegawa Y, Loidl R, Badurek C, Baron M, Manini N, Pistolesi
F and Rauch H 2002 {\em Phys.\ Rev.} A {\bf 65} 052111
\item Hasegawa Y, Zavisky M, Rauch H and Joffe A I 1996 {\em
Phys.\ Rev.\ A} {\bf 53} 2486
\item Gerry C G 1989 {\em Phys.\ Rev.\ A} {\bf 39} 3204
\item Cerver\'o J M and Lejarreta J D 1998 {\em J. Phys. A: Math.
Gen.} {\bf 31} 5507
\item Cerver\'o J M and Lejarreta J D 1997 {\em Quantum Semiclass.
Opt.} {\bf 9} L5
\item Swensmark H and Dimon P 1994 {\em Phys.\ Rev.\ Lett.} {\bf
73} 3387
\item Berry M V 1987 {\em J.\ Mod.\ Opt.} {\bf 34} 1401
\item Ben-Aryeh Y 2003 {\em J.\ Mod.\ Opt.} {\bf 50} (in press)
\item Giller S, Kosinsky P and Szymanowski L 1989 {\em Int.\ J.\
Mod.\ Phys.} A {\bf 4} 1453
\item Bouchiat C 1989 {\em J.\ Phys.\ France} {\bf 50} 1041
\item Joshi A, Pati A K and Banerjee A 1994 {\em Phys.\ Rev.} A {\bf 49}
5131
\item Khanna G, Mukhopadhyay S, Simon R and Mukunda N 1997 {\em
Ann.\ Phys.\ (NY)} {\bf 253} 55
\item Bukanov I V, Tolkachev E A and Tregubovich A Y 1996 {\em
Phys.\ Atom.\ Nucl.} {\bf 59} 659
\item Arvind, Mallesh K S and Mukunda N 1997 {\em J.\ Phys.\ A:
Math.\ Gen.} {\bf 30} 2417
\item Cheng C M and Fung P C W 1989 {\em J.\ Phys.\ A: Math.\
Gen.} {\bf 22} 3493
\item Chaturvedi S, Sriram M S and Srinivasan V 1987 {\em J.\
Phys.\ A: Math.\ Gen.} {\bf 20} L1071
\item Seshadri S, Lakshmibala S and Balakrishnan V 1997 {\em
Phys.\ Rev.} a {\bf 55} 869
\item Bouchiat C and Gibbons G W 1988 {\em J.\ Phys.\ France} {\bf
49} 187
\item Olshani M A 1994 {\em Phys.\ Lett. A}  {\bf 186} 369
\item Lawande Q V, Lawande S V and Joshi A 1999 {\em Phys.\ Lett.\
A} {\bf 251} 164 
\item Moore D J and Stedman G E 1990 {\em J. Phys.\ A: Math.\
Gen.} {\bf 23} 2049
\item Reich M, Sterr U and Ertmer W 1993 {\em Phys.\ Rev.\ A} {\bf
47} 2518
\item Ellinas D, Barnett S M and Dupertuis M A 1989 {\em Phys.\
Rev. A} {\bf 39} 3228
\item Ortega R and Santander M 2003 {\em J.\ Phys.\ A: Math.\
Gen.} {\bf 36} 459
\item Abdel-Aty M 2003 {\em J.\ Opt.\ B: Quantum Semiclass.\ Opt.}
{\bf 5} 349
\item Giavarini G and Onofri E 1989 {\em J.\ Math.\ Phys.} {\bf
30} 659. \item Klyshko D N 1989 {\em Phys.\ Lett.} A {\bf 140} 19
\item Wang X-B, Kwek L C and Oh C H 2000 {\em Phys.\ Rev.\ A} {\bf
62} 032105
\item Liang M-L and Wu H-B 2003 {\em Physica Scripta} T {\bf 68}
41
\item Maamache M and Bekkar H 2003 {\em J.\ Phys.\ A: Math.\ Gen}
{\bf 36} L359 
\item Moore D J 1992 {\em Quant.\ Opt.} {\bf 4} 123
\item Bose S, Carolo A, Fuentes-Guridi I, Santos M F and Vedral V
2003 {\em J.\ Mod.\ Opt.} {\bf 50} 1175
\item Xiang-Bin W and Keiji M 2001 {\em Phys.\ Rev.\ Lett.} {\bf
87} 097901
\item Fuentes-Guridi I, Bose S and Vedral V 2000 {\em Phys.\ Rev.\
Lett.} {\bf 85} 5018
\item Ge Y C and Child M S 1997 {\em Phys.\ Rev.\ Lett.} {\bf 78}
2507
\item Asorey M, Carinena J F and Paramio M 1982 {\em J.\ Math.\
Phys.} {\bf 23} 1451
\item Wu Y-S and Li H-Z 1988 {\em Phys.\ Rev.\ B} {\bf 38} 11907
\item Pati A K 1992 {\em J.\ Phys.\ A: Math.\ Gen.} {\bf 25}
L1001
\item Pati A K 1999 {\em Phys.\ Rev.\ A} {\bf 60} 121
\item Cen L, Li X, Yan Y, Zheng H and Wang S 2003 {\em Phys.\
Rev.\ Lett.} {\bf 90} 147902
\item Yang L and Yan F 2000 {\em Phys.\ Lett.\ A} {\bf 265} 326
\item Anandan J 1990 {\em Phys.\ Lett. A} {\bf 147} 3 
\item Bohm A, Boya L J, Mostafazadeh and Rudolph G 1993 {\em J.\
Geom.\ Phys.} {\bf 12} 13
\item Sj\"oqvist E 2001 {\em Phys.\ Lett.} A {\bf 286} 4
\item Kritsis E 1987 {\em Common.\ Math.\ Phys.} {\bf 111} 417
\item Boya L J, Perelomov A M and Santander M 2001 {\em J.\ Math.\
Phys.} {\bf 42} 5130
\item Jordan T F 1988 {\em J.\ Math.\ Phys.} {\bf 29} 2042
\item Pati a K 2003 {\em Int.\ J.\ Quantum Inf.} {\bf 1} 135
\item Sj\"oqvist E, Pati A K, Ekert A, Anandan J S, Ericson M, Oi
D K L and Vedral V 2000 {\em Phys.\ Rev.\ Lett.} {\bf 85} 2845
\item Tidstrom J and Sj\"qvist E 2003 {\em Phys.\ Rev.} A {\bf 67}
032110
\item Larsson P and Sj\"oqvist E 2003 {\em Phys.\ Lett. A}
{\bf 315} 12
\item Pati A K 1991 {\em Phys.\ Lett.} A {\bf 159} 105.
\item Segert J 1987 {\em J.\ Math.\ Phys.} {\bf 28} 2102.
\item Andreev A V, Klimov A B and Lerner P B 1990 {\em Europhys.\
Lett.} {\bf 12} 101
\item Moore D J 1990 {\em J.\ Phys.\ A: Math.\ Gen} {\bf 23}
L665
\item Byrd M 1998 {\em J.\ Math.\ Phys.} {\bf 39} 6125 
\item Segev M, Solomon R and Yariv A 1992 {\em Phys.\ Rev.\ Lett.}
{\bf 69} 590
\item Tian M, Reibel R R, Barber Z B, Fischer J A and Babbit W R
2003 {\em Phys.\ Rev. A} {\bf 67} 011403(R)
\item Weinfurther H and Badurek G 1990 {\em Phys.\ Rev.\ Lett.}
{\bf 64} 1318
\item Tycko R 1987 {\em Phys.\ Rev.\ Lett.} {\bf 58} 2281
\item  Webb C L, Godun R M, Summy G S, Oberthaler M K, Featonby P
D, Foot C J and Burnett K 1999 {\em Phys.\ Rev.} A {\bf 60} R1783.
\item Richardson D J, Kilvington A I, Green K and Lamoreaux S K
1988 {\em Phys.\ Rev. Lett.} {\bf 61} 2030.
\item Tompkin W R,
Malcuit M S and Boyd R W and Chiao R Y 1990 {\em J.\ Opt.\ Soc.
Am.\ B} {\bf 7} 230.
\item Vedral V 2003 {\em Int. J. Quantum Inf.} {\bf 1} 1.

\end{enumerate}
\end{document}